\documentclass[12pt, draftclsnofoot, onecolumn,comsoc]{IEEEtran}
\usepackage{graphicx}
\usepackage{amsmath,bbm,epsfig,amssymb,amsfonts,amstext,verbatim,amsopn,cite,subfigure,multirow,multicol,lipsum}
\usepackage{balance}
\usepackage{url}
\usepackage{amsfonts}
\usepackage{epsfig}
\usepackage{setspace}
\usepackage{stmaryrd}
\usepackage{psfrag}	
\usepackage{multirow}
\usepackage{float}
\usepackage[process=auto]{pstool}
\usepackage{etoolbox}
\usepackage{algorithm}
\usepackage{algorithmic}
\allowdisplaybreaks

\newtheorem{remk}{Remark}

\newtoggle{OneColumn}

\toggletrue{OneColumn}

\iftoggle{OneColumn}{%
  
}{%
}

\EndPreamble
\begin{document}

\title{Symbol Synchronization for Diffusion-Based Molecular Communications \vspace{-0.1cm}}

\author{Vahid Jamali, Arman Ahmadzadeh, and Robert Schober  \vspace{-1.5cm}
\thanks{V. Jamali, A. Ahmadzadeh, and R. Schober are with the Institute for Digital Communications,
Friedrich-Alexander University (FAU), Erlangen D-91058, Germany (e-mail:
vahid.jamali@fau.de; arman.ahmadzadeh@fau.de; robert.schober@fau.de).}
\thanks{This paper has been presented in part at IEEE ICC 2017 \cite{ICC2017_MC_IEEE}.}  
}

\maketitle

\begin{abstract}
Symbol synchronization refers to the estimation of  the start of a symbol interval and is needed for reliable detection. In this paper, we develop several symbol synchronization schemes for molecular communication (MC) systems where we consider some practical challenges which have not been addressed in the literature yet. In particular, we take into account that in MC systems, the transmitter may not be equipped with an internal clock and may not be able to emit molecules with a fixed release frequency. Such restrictions hold for practical nanotransmitters, e.g. modified cells, where the lengths of the symbol intervals may vary due to the inherent randomness in the availability of food and energy for molecule generation, the process for molecule production, and the release process.  
To address this issue, we develop two synchronization-detection frameworks which both employ two types of molecule. In the first framework, one type of molecule is used for symbol synchronization and the other one is used for data detection, whereas in the second framework, both types of molecule are used for joint symbol synchronization and data detection. For both frameworks, we first derive the optimal maximum likelihood (ML) symbol synchronization schemes as performance upper bounds. Since ML synchronization entails high complexity,  for each framework, we also propose three low-complexity suboptimal schemes, namely a linear filter-based scheme, a peak observation-based scheme, and a threshold-trigger scheme  which are suitable for MC systems with limited computational capabilities. Furthermore, we study the relative complexity and the constraints associated with the proposed schemes and the impact of the insertion and deletion errors that arise due to imperfect synchronization. Our simulation results reveal the effectiveness of the proposed synchronization~schemes and suggest that the end-to-end performance of MC systems significantly depends on the accuracy of the symbol~synchronization. 
\end{abstract}

\begin{IEEEkeywords} 
Diffusive molecular communications, symbol synchronization, and insertion and deletion errors.
\end{IEEEkeywords} 

\section{Introduction}

Recent advances in biology, nanotechnology, and medicine have given rise to the need for communication between nano/micrometer scale nodes \cite{Nariman_Survey,CellBio}.  Employing molecules as information carriers, molecular communication (MC) has quickly emerged as a bio-inspired approach for synthetic communication in micro/nanoscale networks. In fact, calcium signaling among neighboring cells, the use of neurotransmitters for communication across the synaptic cleft of neurons, and the exchange of autoinducers by bacteria for quorum sensing are among the many examples of MC in nature \cite{CellBio}.

\subsection{Prior Work on Synchronization in MC}

One of the crucial requirements for establishing a reliable communication link is symbol synchronization where the start of a symbol interval is determined at the receiver.  Most  works available in the literature on MC assume perfect symbol synchronization for data detection, see e.g. \cite{ArmanMobileMC,Chae_Absorbing,MoSK_Yilmaz,NanoCOM16}. First studies on establishing a synchronization mechanism for MC systems have been conducted in  \cite{MC_Akyildiz_Sync,MC_Nakano_Sync,MC_Sync_Bio,MC_Clock_Sync,MC_Gaussian_Sync,Clock_Drift_MC,OffsetSync_MC,MC_Blind_Sync}. In particular, in \cite{MC_Akyildiz_Sync,MC_Nakano_Sync,MC_Sync_Bio}, the authors proposed a  scheme for synchronizing multiple molecular machines that have to carry out a common task, e.g., coordinate their behavior as in quorum sensing among bacteria. \textit{Symbol synchronization} was investigated in \cite{MC_Clock_Sync,MC_Gaussian_Sync,Clock_Drift_MC,OffsetSync_MC,MC_Blind_Sync}. In \cite{MC_Clock_Sync,MC_Gaussian_Sync,Clock_Drift_MC,OffsetSync_MC}, the authors proposed a two-way message exchange protocol between the transmitter and the receiver facilitating the estimation and correction of \textit{constant} frequency and delay offsets between the clocks of the transmitter and the receiver. However, to achieve high performance, the synchronization protocols in \cite{MC_Clock_Sync,MC_Gaussian_Sync,Clock_Drift_MC,OffsetSync_MC} require several rounds of two-way message exchange between the transmitter and the receiver which leads to a large overhead considering the slow propagation of molecules in  MC channels. Furthermore, for cases when flow is present in the environment, e.g. in the direction from the transmitter to the receiver, it may not be possible to establish a feedback link  from the receiver to the transmitter. To reduce the synchronization overhead, the authors in \cite{MC_Blind_Sync} proposed a blind synchronization scheme based on a sequence of data molecules observed at the receiver. However, in \cite{MC_Blind_Sync}, the clocks of the transmitter and the receiver are assumed to have identical frequencies and only a \textit{constant} clock offset may exist.

\subsection{Our Contributions}

In this paper, we develop two new symbol synchronization frameworks that take into account some practical challenges of MC systems which have not been considered in \cite{MC_Clock_Sync,MC_Gaussian_Sync,Clock_Drift_MC,OffsetSync_MC,MC_Blind_Sync}. In particular, in \cite{MC_Clock_Sync,MC_Gaussian_Sync,Clock_Drift_MC,OffsetSync_MC,MC_Blind_Sync}, similar to wireless communications \cite{Clock_Sync}, it is assumed  that the nodes are equipped with internal clocks and accurate oscillators. Thereby, the problem of synchronization was reduced to the elimination of possible frequency and delay offsets between the clocks. Furthermore, in \cite{MC_Clock_Sync,MC_Gaussian_Sync,Clock_Drift_MC,OffsetSync_MC,MC_Blind_Sync}, it is assumed that the transmitter emits molecules  with a fixed release frequency, i.e., the  symbol durations are constant and identical.  However, in a real MC system, the transmitter will be a biological or electronic nanomachine, e.g. a modified cell, which controls the release of the information molecules into the channel using e.g. electrical, chemical, or optical signals \cite{CellBio,Hamid_Ion_Modul}. Because of the inherent randomness in the availability of food and energy for molecule generation, the process for molecule production, and the release process, see \cite[Chapters~12 and 13]{CellBio},  in practical MC systems, the lengths of the symbol intervals may vary. 

To cope with the aforementioned practical challenges, in this paper, we develop symbol synchronization schemes for MC systems where the transmitter is not required to be equipped with an internal clock nor restricted to release the molecules with a constant frequency. To facilitate simultaneous symbol synchronization and data detection, we employ two types of molecule, and propose two different synchronization-detection frameworks: \textit{i) Framework~1 (independent symbol synchronization and data detection):} In this framework, one type of molecule is used for symbol synchronization and the other one is used for data detection. \textit{ii) Framework~2 (joint symbol synchronization and data detection):} Here, both types of molecule are used for joint symbol synchronization and data detection. Both frameworks entail different degrees of synchronization-detection accuracy, complexity, and applicability as will be discussed in detail throughout the paper.  For both proposed frameworks, we first derive the optimal maximum likelihood (ML) symbol synchronization scheme as performance upper bound. Since ML synchronization entails high complexity, for each framework, we also propose three suboptimal schemes, namely a linear filter-based (LF) scheme, a peak observation-based (PO) scheme, and a  threshold-trigger (TT) scheme, which are suitable for MC systems with limited computational capabilities. We further discuss the advantages and disadvantages of the proposed synchronization schemes in terms of their required a priori knowledge, relative complexity, and applicability. In addition, we study insertion and deletion errors, which arise due to imperfect synchronization. We further apply an error-correction code from \cite{InsDelMarkerCode} to mitigate these errors. Our simulation results unveil the effectiveness of the proposed synchronization~schemes and suggest that the end-to-end performance of MC systems significantly depends on the accuracy of the symbol~synchronization.

We note that this paper expands its conference version \cite{ICC2017_MC_IEEE} in several directions. First, in \cite{ICC2017_MC_IEEE}, we  studied only  independent symbol synchronization and data detection, i.e., Framework~1. Second, the ML symbol synchronization problem formulated in this paper is different from the one in \cite{ICC2017_MC_IEEE}. Third, the proposed LF symbol synchronization scheme was not considered in \cite{ICC2017_MC_IEEE}. Finally, many of the extensive discussions and simulation results are not included in \cite{ICC2017_MC_IEEE}. 

The remainder of this paper is organized as follows.
In Section II, the considered system and signal models are presented. The proposed synchronization and detection schemes are introduced in Section~III, and their properties are discussed in more detail in Section~IV. Numerical results are reported in
Section V, and conclusions are drawn in Section VI.

\section{System and Signal Models}

In this section, we first present the  MC system model considered in this paper. Subsequently, we introduce the signal models used for synchronization and data transmission.

\subsection{System Model}

\begin{figure}
  \centering
 \scalebox{1}{
\pstool[width=0.9\linewidth]{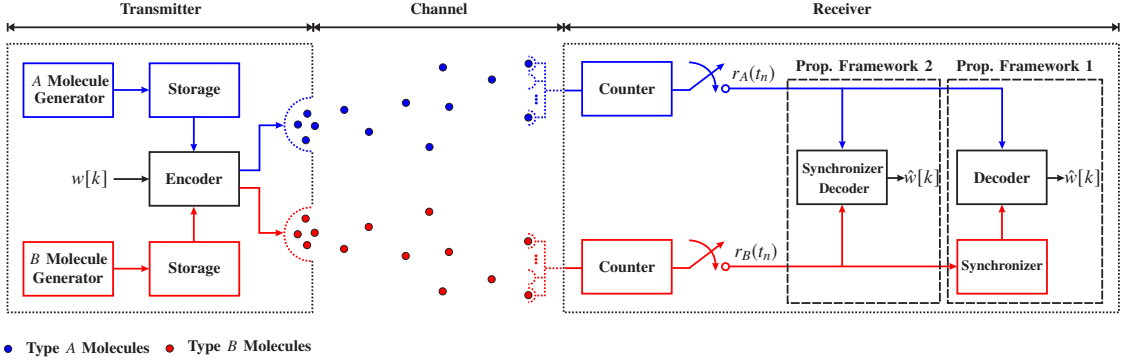}{
\psfrag{AA}[c][c][0.5]{\textbf{Generator}}
\psfrag{AB}[c][c][0.5]{$A$ \textbf{Molecule}}
\psfrag{BA}[c][c][0.5]{$B$ \textbf{Molecule}}
\psfrag{BB}[c][c][0.5]{\textbf{Storage}}
\psfrag{CC}[c][c][0.5]{\textbf{Encoder}}
\psfrag{DD}[c][l][0.6]{$w[k]$}
\psfrag{EE}[c][c][0.5]{\textbf{Counter}}
\psfrag{HH}[c][r][0.6]{$r_A(t_n)$}
\psfrag{GG}[c][r][0.6]{$r_B(t_n)$}
\psfrag{F}[c][c][0.45]{\textbf{Synchronizer}}
\psfrag{JJ}[c][r][0.6]{$\hat{w}[k]$}
\psfrag{II}[c][c][0.5]{\textbf{Decoder}}
\psfrag{Sd1}[c][c][0.45]{\textbf{Synchronizer}}
\psfrag{Sd2}[c][c][0.45]{\textbf{Decoder}}
\psfrag{KK}[c][c][0.5]{\textbf{Transmitter}}
\psfrag{LL}[c][c][0.5]{\textbf{Channel}}
\psfrag{MM}[c][c][0.5]{\textbf{Receiver}}
\psfrag{D}[l][c][0.5]{\textbf{Type $A$ Molecules}}
\psfrag{S}[l][c][0.5]{\textbf{Type $B$ Molecules}}
\psfrag{F1}[c][c][0.5]{\textbf{Prop. Framework 2}}
\psfrag{F2}[c][c][0.5]{\textbf{Prop. Framework 1}}
} }  \vspace{-0.01cm}
\caption{Block diagram of the considered MC setup for both Framework~1, i.e., independent synchronization (using type-$B$ molecules) and detection (using type-$A$ molecules), and Framework~2, i.e., joint synchronization and detection (using both type-$A$ and type-$B$ molecules). \vspace{-0.03cm}}
\label{Fig:Block}
\end{figure}

 We consider an MC system  consisting of a transmitter, a channel, and a receiver, see Fig.~\ref{Fig:Block}. The transmitter is able to release two types of molecules, namely type-$A$ and type-$B$ molecules which facilitates simultaneous symbol synchronization and data detection. Exploiting these two types of molecule, we consider the following two synchronization and detection frameworks: 
\begin{itemize}
\item[\textit{i)}] \textit{Framework~1 (Independent Synchronization and Detection):} Type-$B$ molecules are employed  for synchronization and type-$A$ molecules are used for information transmission.
\item[\textit{ii)}] \textit{Framework~2 (Joint Synchronization and Detection):} Type-$A$ and type-$B$ molecules are used for joint information transmission and synchronization.
\end{itemize}
The advantageous and  disadvantageous of the above two frameworks are discussed in detail in Section~IV-A. For instance, joint ML synchronization-detection under Framework~2 outperforms the independent ML synchronization and ML detection under Framework~1 in terms of the end-to-end bit error rate (BER). However, Framework~1 offers more flexibility in the sense that any arbitrary modulation and detection schemes can be employed for data transmission independent of the adopted symbol synchronization scheme.

The details of how the molecules are released by the transmitter for the above two frameworks will be explained in the next subsection.
The released molecules diffuse through the fluid medium between the transmitter and the receiver.  The movements of individual molecules are assumed to be independent from each other. Furthermore, we assume that the molecules of types $A$ and $B$ have identical diffusion coefficients denoted by $D$ \cite{Nariman_Survey}. We consider a spherical receiver whose surface is partially covered by two different types of receptors for detecting type-$A$ and type-$B$ molecules, respectively  \cite{Arman_ReactReciever}. Molecules that reach the receiver can participate in a reversible bimolecular reaction with receiver receptor proteins. Thereby, the receiver treats the time-varying numbers of type-$A$ and type-$B$ molecules bound to the receptors as the received signals for data detection and synchronization.  

The MC channel is characterized by the following two quantities. \textit{i)} The \textit{expected} number of type-$x$ molecules bound to the corresponding receptors at the receiver at time $t$ due to the release of molecules by the transmitter in \textit{one} symbol interval starting at $t=0$, which is denoted by $P_x(t),\,\,x\in\{A,B\}$. \textit{ii)} The \textit{expected} number of external noise molecules bound to the receptors, denoted by $z_x,\,\,x\in\{A,B\}$. In general, $P_x(t),\,\,x\in\{A,B\}$,  depends on the release mechanism at the transmitter, the MC environment, and the properties of the receiver such as its size, the number of receptors, etc. For instance, assuming instantaneous molecule release and a point source transmitter,   expressions for $P_x(t)$ can be found in \cite{Arman_ReactReciever} for a general reactive receiver and in \cite{Chae_Absorbing} for an absorbing receiver. On the other hand, the external noise molecules originate from other MC links  or natural sources which also employ type-$A$ or type-$B$ molecules. We emphasize that the synchronization and detection schemes proposed in this paper are general and are applicable for any given expression for $P_x(t)$ and any value of $z_x$. For future reference, we refer to ${\mathrm{SNR}}_x=\frac{{\mathrm{max}}_{t\geq 0}\,\, P_x(t)}{z_x},\,\,x\in\{A,B\}$, as the signal-to-noise ratio (SNR) for type-$x$ molecules.

\subsection{Signal Model}

Let  $w[k]\in\{0,1\}$ denote the binary data symbol in the $k$-th symbol interval. We assume $\mathrm{Pr}\{w[k]=1\}=\mathrm{Pr}\{w[k]=0\}=0.5$ where $\mathrm{Pr}\{X\}$ denotes the probability of event $X$.  The transmitter wishes to continuously send data symbols; however, the release time of the molecules at the transmitter may vary from one symbol interval to the next due to variations in the availability of food and energy for molecule generation, the rate for molecule production, and the release process  over time, see \cite[Chapters~12 and 13]{CellBio}. To model the aforementioned effects, let $t_s[k]\in\mathcal{T}[k]$ denote a random variable (RV) whose realization specifies the start of the $k$-th symbol interval where $\mathcal{T}[k]$ is given by
\begin{IEEEeqnarray}{lll} \label{Eq:ML_Tk}
\mathcal{T}[k]=[t_s[k-1]+T^{\min},t_s[k-1]+T^{\max}].
\end{IEEEeqnarray}
The duration of the $k$-th symbol interval is the time elapsed between $t_s[k]$ and $t_{s}[k+1]$; hence, in (\ref{Eq:ML_Tk}), $T^{\min}$ and $T^{\max}$ are in fact the  minimum and maximum possible lengths of a symbol interval, respectively. In other words, the length of each symbol interval is an RV in $[T^{\min},T^{\max}]$. Note that the symbol rate of the considered MC system, denoted by $R$, is bounded by $\frac{1}{T^{\max}}\leq R \leq \frac{1}{T^{\min}}$.

Let $a[k]$ be a binary  variable which is equal to one if type-$A$ molecules are released at the beginning of the $k$-th symbol interval, and equal to zero otherwise. Similarly, let $b[k]$ denote a binary  variable which is equal to one if type-$B$ molecules are released at the beginning of the $k$-th symbol interval, and equal to zero otherwise. In the following, we specify $a[k]$ and $b[k]$ for the two considered transmission frameworks. 
\begin{itemize}
\item[\textit{i)}] \textit{Framework~1:}   To establish symbol synchronization, at the beginning of \textit{each} symbol interval, the transmitter releases $N_B$ type-$B$ molecules, i.e., $b[k]=1,\,\,\forall k$. Moreover, depending on whether $w[k]=1$ or $w[k]=0$ holds, the transmitter releases either $N_A$ or zero type-$A$ molecules, respectively, i.e., $a[k]=w[k],\,\,\forall k$~\footnote{In practical MC systems, the number of  molecules released by the transmitter may not be constant and may also vary from one symbol interval to the next. For simplicity, in this paper, we assume that the transmitter waits until a sufficient number of molecules is available and then releases exactly $N_B$ synchronization and/or $N_A$ information molecules, respectively. The extension of the proposed synchronization schemes  to account for varying numbers of released molecules is an interesting topic for future research.}. In other words, ON-OFF keying modulation is performed~\cite{Nariman_Survey}. 

\item[\textit{ii)}] \textit{Framework~2:}  Here, we employ type-$A$ and type-$B$ molecules for joint synchronization and data transmission. In particular, depending on whether $w[k]=1$ or $w[k]=0$ holds, the transmitter releases either $N_A$ type-$A$ molecules or $N_B$ type-$B$ molecules, respectively, i.e., $a[k]=w[k]$ and $b[k]=1-w[k],\,\,\forall k$. We note that using different types of moelcules for modulation has already been  proposed in \cite{MoSK_Yilmaz} and is referred to as molecule shift keying (MoSK) modulation. However, for MoSK,  perfect synchronization is assumed in \cite{MoSK_Yilmaz}, whereas under the proposed Framework~2, we aim to develop a joint synchronization and detection~scheme.
\end{itemize}

To model the received signal, we assume that the receiver periodically counts the numbers of type-$A$ and type-$B$ molecules bound to the respective receptors on its surface with a frequency of $\Delta t$ seconds. Therefore, time can be discretized into a sequence of observation time samples $t_n=(n-1)\Delta t,\,\,n=1,2,\dots$, at the receiver. Moreover, let us define  $r_x(t_n)$ as the number of type-$x$ molecules bound to the respective receptors  at sample time $t_n$. Since at any given time after the release of the molecules by the transmitter, the molecules are either bound to a receptor or not, a binary state model applies and the number of bound molecules follows a binomial distribution. We note that the binomial distribution converges to the Poisson distribution when the number of trials is high and the success probability is small \cite{Adam_Enzyme}. These assumptions are justified for MC since the number of released molecules is typically very large
and the probability that any given molecule released by the transmitter reaches the receiver is typically very small \cite{TCOM_MC_CSI}. Therefore,  $r_x(t_n)$ can be modeled as follows \cite{Yilmaz_Poiss,HamidJSAC,Cl_MF_IEEE}
\begin{IEEEeqnarray}{lll} \label{Eq:InputOutput}
  r_x(t_n)  =  \mathcal{P}(\bar{r}_x(t_n)), \quad x\in\{A,B\},
\end{IEEEeqnarray}
where $\mathcal{P}(\lambda)$  denotes a Poisson RV with mean  $\lambda$ and $\bar{r}_x(t_n)$ is the mean number of type-$x$ molecules bound to the receiver's receptors at time $t_n$, i.e., $\bar{r}_x(t_n)=\mathcal{E}\{r_x(t_n)\}$ where $\mathcal{E}\{\cdot\}$ denotes expectation. Hence, we obtain
\begin{IEEEeqnarray}{lll} \label{Eq:ExpMol}
  \bar{r}_A(t_n)  &=  \sum_{\forall k|t_s[k]\leq t_n}  a[k] P_A\left(t_n- t_s[k]\right) + z_A, \IEEEyesnumber\IEEEyessubnumber\\
    \bar{r}_B(t_n) & =  \sum_{\forall k|t_s[k]\leq t_n}  b[k] P_B\left(t_n- t_s[k]\right) + z_B. \IEEEyessubnumber
\end{IEEEeqnarray}

\section{Proposed Synchronization and Detection Schemes}
In this section, we develop several synchronization and detection schemes for the proposed frameworks.

\subsection{Independent Synchronization and Detection  (Framework~1)}

 In the following, we first develop optimal and suboptimal synchronization schemes. Subsequently, we present the adopted detection scheme. 
  
\subsubsection{Optimal ML Scheme}
Our goal is to determine the start of each symbol interval, i.e., $t_s[k]$, based on the received signal for type-$B$ molecules, i.e., $r_B(t_n),\,\,\forall t_n$.  \textit{Joint ML symbol} synchronization of several consecutive symbol intervals entails a very high computational complexity due to the multi-dimensional nature of the corresponding ML hypothesis test. Therefore, we focus on the formulation of an ML problem for \textit{symbol-by-symbol} synchronization which is computationally tractable. To this end, we introduce two assumptions which enable us to formulate an ML problem for estimating $t_s[k]$ without knowledge of $t_s[k'],\,\,k'>k$. Before presenting these assumptions, let us first define $\mathcal{T}^{\mathrm{ow}}[k]$ as the set of the observation samples used to compute the ML metric for each hypothesis time $t$ for $t_s[k]$, i.e.,  observation samples $t_n \in \mathcal{T}^{\mathrm{ow}}[k]$ are used for hypothesis test $t$.
 
\begin{figure}
  \centering
 \scalebox{0.8}{
\pstool[width=0.75\linewidth]{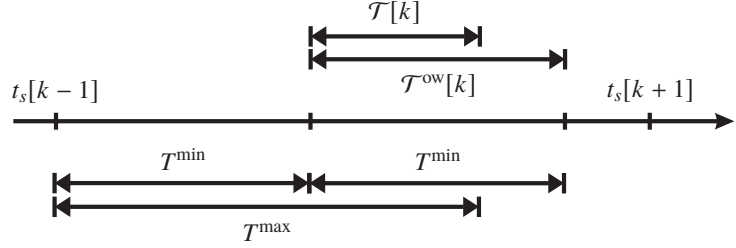}{
\psfrag{t1}[c][c][1]{$t_s[k-1]$}
\psfrag{ts}[c][c][1]{$t$}
\psfrag{t2}[c][c][1]{$t_s[k+1]$}
\psfrag{tmax}[c][c][1]{$T^{\max}$}
\psfrag{tmin}[c][c][1]{$T^{\min}$}
\psfrag{to}[c][c][1]{$\mathcal{T}^{\mathrm{ow}}[k]$}
\psfrag{tk}[c][c][1]{$\mathcal{T}[k]$}
} } \vspace{-0.3cm} 
\caption{Illustration of assumptions A1, i.e., $t_s[k+1]\notin\mathcal{T}[k]$, and A2, i.e., $t_s[k+1]\notin\mathcal{T}^{\mathrm{ow}}[k]$, adopted for development of the symbol-by-symbol ML synchronization problem. \vspace{-0.2cm}}
\label{Fig:MLProb}
\end{figure}

\begin{itemize}
\item[A1:] We assume that $t_s[k+1]\notin\mathcal{T}[k]$ holds which leads to the condition $T^{\max}\leq 2T^{\min}$. We note that if $t_s[k+1]\in\mathcal{T}[k]$ can  occur, $t=t_s[k+1]$ may be selected as the ML estimate for the $k$-th symbol interval.
\item[A2:] We assume that $t_s[k+1]\notin\mathcal{T}^{\mathrm{ow}}[k]$ holds which ensures that the ML metric for $t$ is not affected by the value of $t_s[k+1]$. Note that the observation samples $r_B(t_n)$ at $t_n<t_s[k-1]+T^{\min}$ are not affected by $t_s[k]$. Therefore, the largest feasible observation set which ensures $t_s[k+1]\notin\mathcal{T}^{\mathrm{ow}}[k]$ is $\mathcal{T}^{\mathrm{ow}}[k]=[t_s[k-1]+T^{\min},\,t_s[k-1]+2T^{\min}]$. 
\end{itemize}

The above assumptions are schematically illustrated in Fig.~\ref{Fig:MLProb}. Based on assumptions A1 and A2, the ML problem can be mathematically formulated as 
\begin{IEEEeqnarray}{rl} \label{Eq:ML_Sync}
\hat{t}_s^{\mathrm{ml}}[k] \,&= \underset{\forall t \in\mathcal{T}[k] }{\mathrm{argmax}}\,\, \Lambda_B^{\mathrm{ml}}(t)  \\
\Lambda_B^{\mathrm{ml}}(t) \,&\triangleq\prod_{t_n\in\mathcal{T}^{\mathrm{ow}}[k]}  f_{\mathcal{P}}\big(r_B(t_n), \bar{r}_B(t_n)|{t_s[k]=t}\big),\label{Eq:ML_Sync_metric}
\end{IEEEeqnarray}
where $f_{\mathcal{P}}(x,\lambda)=\frac{\lambda^xe^{-\lambda}}{x!}$ is the probability mass function of a Poisson RV with mean $\lambda$.
In (\ref{Eq:ML_Sync}), it is assumed that the observations  $r_B(t_n)$ at different time instants are independent such that the likelihood function over observation window $t_n\in\mathcal{T}^{\mathrm{ow}}[k]$ can be expressed as the product of the likelihood functions for each time instant $t_n$, $f_{\mathcal{P}}\big(r_B(t_n), \bar{r}_B(t_n)|{t_s[k]=t}\big)$. Moreover, for a given hypothesis $t$ for $t_s[k]$, assuming that the  ML estimate of symbol interval $k'<k$ was correct,   $\bar{r}_B(t_n)$ is given by (\ref{Eq:ExpMol}b).

 Maximizing $\Lambda_B^{\mathrm{ml}}(t)$ is equivalent to maximizing $\mathrm{ln}(\Lambda_B^{\mathrm{ml}}(t))$ since $\mathrm{ln}(\cdot)$ is a monotonically increasing function. Hence, the ML problem in (\ref{Eq:ML_Sync}) can be rewritten as
\begin{IEEEeqnarray}{ll} \label{Eq:ML_log}
\hat{t}_s^{\mathrm{ml}}[k] &= \underset{\forall t \in\mathcal{T}[k] }{\mathrm{argmax}}\,\,\mathrm{ln}(\Lambda_B^{\mathrm{ml}}(t))  \nonumber\\
&=\underset{\forall t \in\mathcal{T}[k] }{\mathrm{argmax}}\sum_{t_n\in\mathcal{T}^{\mathrm{ow}}[k]}  \big[ r_B(t_n)\mathrm{ln}(\bar{r}_B(t_n))-\bar{r}_B(t_n)-\mathrm{ln}(r_B(t_n)!) \big].
\end{IEEEeqnarray}
Although the problem in  (\ref{Eq:ML_log}) does not lend itself to an elegant closed-form solution, we can find the optimal ML solution numerically using a simple one-dimensional search. 

\begin{remk}
The proposed ML synchronization scheme provides optimal symbol synchronization at the cost of a high computational complexity which may not be affordable for implementation at nanoscale. Nevertheless, ML synchronization can serve as a benchmark for the low-complexity synchronization schemes proposed in this paper. Moreover, for applications where the nanoreceiver only collects observations, i.e., $r_B(t_n)$, and forwards them to an external processing unit outside the MC environment, the computational complexity of ML synchronization may be affordable. This case may apply e.g. in health monitoring where a computer outside the body may be available for offline processing.
\end{remk}

Fig.~\ref{Fig:Concept} illustrates an example scenario for the proposed  synchronization schemes, where five consecutive symbol intervals are considered where the symbols are chosen as $w[k]=[1 ,1 ,0, 0 ,1]$ and the starts of the symbol intervals are chosen as $t_s[k]=[0, 2.2, 4, 6, 8.4]$ ms, i.e., the transmitter does not release the molecules at a fixed frequency. We refer to this example as \textit{example scenario} and use it to illustrate how the proposed synchronization schemes work throughout the paper. We show the results for Frameworks~1 and 2 on the left-hand side (LHS) and the right-hand side (RHS) of Fig.~\ref{Fig:Concept}, respectively. The LHSs of Figs.~\ref{Fig:Concept}~a) and b) show  for one realization the numbers of received type-$A$ and type-$B$ molecules, respectively, under Framework~1. As can be seen from the LHS of Fig.~\ref{Fig:Concept} c), the proposed ML synchronization scheme is able to accurately determine the start times of the symbol intervals for the  set of parameters considered in  this figure.

 \subsubsection{Suboptimal Low-Complexity Schemes}
  In the following, we propose three suboptimal low-complexity synchronization schemes which may be suitable for implementation in simple nanoreceivers.

\textit{Linear Filter-Based  Scheme:}
Recall that the ML synchronization scheme optimally takes into account all samples within the observation window $t_n\in\mathcal{T}^{\mathrm{ow}}[k]$ for each  possible hypothesis $t\in\mathcal{T}[k]$ in order to estimate $t_s[k]$. To reduce the complexity, we employ a linear filter to derive a metric for estimation of the start of the symbol intervals. In particular, we adopt the expected mean of the received signal as the impulse response of the linear filter which leads to the following LF symbol synchronization scheme
\begin{IEEEeqnarray}{rl} \label{Eq:FilterProt}
\hat{t}^{\mathrm{lf}}_s[k]\,&= \underset{\forall t \in\mathcal{T}[k] }{\mathrm{argmax}} \,\,\tilde{r}_B(t)  \\
\tilde{r}_B(t) \,&\triangleq \sum_{t_n\in\mathcal{T}^{\mathrm{ow}}[k]}  r_B(t_n) \bar{r}_B(t_n), \label{Eq:FilterProt_metric}
\end{IEEEeqnarray}
where $\bar{r}_B(t_n)$ is obtained from (\ref{Eq:ExpMol}b) after substituting $t_s[k]=t$. Comparing (\ref{Eq:FilterProt_metric}) and (\ref{Eq:ML_log}), we can observe that the synchronization scheme based on linear filtering in (\ref{Eq:FilterProt}) requires only linear operations which are computationally simpler than the nonlinear operations required for the optimal ML synchronization scheme in (\ref{Eq:ML_log}). As can be observed from the LHS of Fig.~\ref{Fig:Concept} d), despite its simplicity compared to the ML scheme,  for the considered example scenario, the LF synchronization scheme can accurately find the start times of the symbol intervals.

\textit{Peak Observation-Based  Scheme:}
To further reduce the complexity of synchronization, we propose to estimate $t_s[k]$ based on only the peak observation. We note that the estimation of the release times of molecules by a transmitter based on peak observations at the receiver has been proposed in \cite{Adam_Channel} for transmission of only one symbol. In contrast, in this paper, we consider the practical case where multiple consecutive symbols are transmitted, and hence unlike \cite{Adam_Channel}, the estimation accuracy in each symbol interval influences the estimation performance of future symbols.  To formally present the proposed PO synchronization scheme, let us first define constant $t^{\mathrm{p}}={\mathrm{argmax}}_{t\geq 0}\,\,P_B(t)$. Thereby, assuming perfect estimation of $t_s[k'],\,\,\forall k'<k$,  the set of expected time instances where the peak observation of the synchronization molecules in symbol interval $k$ can occur is given by 
\begin{IEEEeqnarray}{lll} \label{Eq:Peak_Tp}
\mathcal{T}^{\mathrm{p}}[k]=\left[t_s[k-1]+T^{\min}+t^{\mathrm{p}},t_s[k-1]+T^{\max}+t^{\mathrm{p}}\right].\quad
\end{IEEEeqnarray}
Hereby, we propose a PO symbol synchronization scheme which estimates the start of the symbol intervals as follows
\begin{IEEEeqnarray}{lll} \label{Eq:PeakProt}
\hat{t}^{\mathrm{po}}_s[k]= \Big[\underset{t_n\in \mathcal{T}^{\mathrm{p}}[k]}{\mathrm{argmax}} \,\,r_B(t_n) \Big] - t^{\mathrm{p}}.\quad
\end{IEEEeqnarray}
On the LHS of Fig.~\ref{Fig:Concept} e), the above PO synchronization is schematically illustrated for the example scenario. While the complexity of the PO synchronization scheme is considerably lower than that of the ML and LF synchronization schemes, as will be shown in detail in Section~V, the corresponding performance loss may be significant.  This motivates us to propose a TT synchronization scheme which is also relatively simple, but provides a better performance compared to PO synchronization scheme.

 \textit{Threshold-Trigger Scheme:}
 In nature, a common strategy among living organisms in response to external stimuli is based on a threshold-trigger mechanism. For example, the increase of the concentration of a certain type of molecule around a cell can trigger a response inside the cell \cite{CellBio}. In the following, we exploit the TT mechanism for symbol synchronization.

The main idea behind our simple TT symbol synchronization scheme is that the receiver considers the number of bound information molecules for detection only while the number of bound synchronization molecules is above a certain threshold. In other words, instead of determining the actual symbol interval, the proposed protocol only determines a detection zone which is used for data detection in each symbol interval.  In order to formally present the proposed scheme, let us define $\xi$ as a constant threshold and $\hat{t}_s^{\mathrm{tt}}[k]$ and $\hat{t}_e^{\mathrm{tt}}[k]$ as the beginning and the end of the detection zone for symbol interval $k$, respectively.  Furthermore, since the number of bound molecules is an RV and may rapidly fluctuate, we assume a minimum detection interval size of $T^{\mathrm{dw}}$ to avoid possible false alarms indicating a new symbol interval. On the other hand, $T^{\mathrm{dw}}\leq T^{\min}$ has to hold to avoid missing the next symbol interval.
We propose the following TT scheme to determine  $\hat{t}_s^{\mathrm{tt}}[k]$ and $\hat{t}_e^{\mathrm{tt}}[k]$:
\begin{IEEEeqnarray}{lll} \label{Eq:SubOpt_Sync}
\hat{t}_s^{\mathrm{tt}}[k]&=\underset{t_n>\hat{t}_e^{\mathrm{tt}}[k-1]}{\mathrm{min}}\,\,t_n|r_B(t_n)\geq \xi  \IEEEyesnumber\IEEEyessubnumber \\
\hat{t}_e^{\mathrm{tt}}[k]&=\mathrm{max}\Big\{\underset{t_n>\hat{t}_s^{\mathrm{tt}}[k]}{\mathrm{min}} t_n| r_B(t_n)\leq \xi, \hat{t}_s^{\mathrm{tt}}[k]+T^{\mathrm{dw}} \Big\}.\,\,\quad \IEEEyessubnumber
\end{IEEEeqnarray}
 In other words, $\hat{t}_s^{\mathrm{tt}}[k]$ in (\ref{Eq:SubOpt_Sync}a) activates detection whereas $\hat{t}_e^{\mathrm{tt}}[k]$ in (\ref{Eq:SubOpt_Sync}b) terminates detection for symbol interval~$k$. 

The LHS of Fig.~\ref{Fig:Concept} f) illustrates the proposed TT synchronization scheme for the example scenario.  As can be seen from this figure, the TT scheme selects many of the observation samples  within a given symbol interval for data detection without explicitly estimating the start times of the symbol interval.

\subsubsection{Detection} 

In order to be able to focus on the effect of imperfect synchronization on the BER performance, we employ optimal ML detection for all (optimal and suboptimal) synchronization schemes proposed for Framework~1. Note that if a suboptimal detector was adopted for the low-complexity suboptimal synchronization scheme, it would be difficult to determine whether the performance loss compared to the ML synchronization scheme is due to imperfect synchronization or suboptimal detection. The ML detector is given by
\begin{IEEEeqnarray}{lll} 
\hat{w}[k] =\begin{cases}
1,\quad &\mathrm{if}\,\,\Lambda_A^{\mathrm{ml}}(w[k]=1) \geq \Lambda_A^{\mathrm{ml}}(w[k]=0)\\
0,\quad &\mathrm{otherwise}
\end{cases} 
\end{IEEEeqnarray}
where $\Lambda_A^{\mathrm{ml}}(w[k])$ is the ML detection metric given by
\begin{IEEEeqnarray}{lll}\label{Eq:ML_Det_Met} 
\Lambda_A^{\mathrm{ml}}(w[k]) \triangleq\prod_{t_n\in\mathcal{T}^{\mathrm{det}}[k]}  f_{\mathcal{P}}\big(r_A(t_n), \bar{r}_A(t_n)|{w[k]}\big).
\end{IEEEeqnarray}
In (\ref{Eq:ML_Det_Met}), $\mathcal{T}^{\mathrm{det}}[k]$ is the detection window which, given the adopted synchronization scheme, is defined as
\begin{IEEEeqnarray}{lll}\label{Eq:Ts_tot} 
\mathcal{T}^{\mathrm{det}}[k]  = \begin{cases}
[\hat{t}_s^{\mathtt{x}}[k],\hat{t}_s^{\mathtt{x}}[k+1]],&\text{ML/LF/PO Sync.}, \,\, \mathtt{x}\in\{\mathrm{ml},\mathrm{lf},\mathrm{po}\}\\
[\hat{t}_s^{\mathrm{tt}}[k],\hat{t}_e^{\mathrm{tt}}[k]],&\text{TT Sync.}
\end{cases}\quad\,\,
\end{IEEEeqnarray}
Note that in (\ref{Eq:ML_Det_Met}), the value of $w[k]$ changes the signal mean $\bar{r}_A(t_n)$, cf. (\ref{Eq:ExpMol}a).  Using the monotonicity property of the logarithm function, the ML detector can be simplified to 
\begin{IEEEeqnarray}{lll} \label{Eq:ML_Det_Log} 
\hat{w}[k] =\begin{cases}
1,\quad &\mathrm{if}\,\,\sum_{t_n\in\mathcal{T}^{\mathrm{det}}[k]} \left[r_A(t_n)\mathrm{ln}(\bar{r}_A(t_n)) - \bar{r}_A(t_n)\right]_{w[k]=1}\\
&\qquad\qquad \geq \sum_{t_n\in\mathcal{T}^{\mathrm{det}}[k]} \left[r_A(t_n)\mathrm{ln}(\bar{r}_A(t_n)) - \bar{r}_A(t_n)\right]_{w[k]=0}\\
0,\quad &\mathrm{otherwise}.
\end{cases} 
\end{IEEEeqnarray}

\subsection{Joint Synchronization and Detection (Framework~2)}

Similar to Framework~1, for Framework~2, we first derive the joint ML synchronization and detection scheme. Then, we present several suboptimal low-complexity schemes for practical MC systems with limited computational capabilities. However, unlike Framework~1, which uses observations $r_A(t_n),\,\forall t_n$, for data detection and observations $r_B(t_n),\,\forall t_n$, for symbol synchronization, Framework~2 employs both observations $r_A(t_n)$ and $r_B(t_n),\,\forall t_n$, for simultaneous symbol synchronization and data detection, cf. Section~II-B.

\subsubsection{Optimal ML Scheme}

In this subsection, our goal is to develop a joint synchronization and detection scheme. In other words, we derive a joint estimator and detector for estimating $t_s[k]$ and $w[k]$, respectively, based on the received signal for type-$A$ and type-$B$ molecules, i.e., $r_A(t_n)$ and $r_B(t_n),\,\,\forall t_n$.  As discussed in Section~III-A, the optimal ML scheme for several consecutive symbol intervals is very complicated due to the multi-dimensional nature of the corresponding ML hypothesis test. Therefore, similar to the design of the optimal ML synchronizer  in Section~III-A, our focus here is to formulate a computationally tractable ML problem for \textit{symbol-by-symbol} joint synchronization and detection. This is possible when assumptions A1 and A2, which were introduced in Section~III-A, hold. In particular, the  ML problem for joint synchronization and detection is formulated as 
\begin{IEEEeqnarray}{ccc} \label{Eq:ML_SyncDet}
[\hat{t}_s^{\mathrm{ml}}[k],\hat{w}^{\mathrm{ml}}[k]] =  \underset{\forall t \in\mathcal{T}[k], w\in\{0,1\} }{\mathrm{argmax}}\,\,
  \Lambda_A^{\mathrm{ml}}(t,w)\Lambda_B^{\mathrm{ml}}(t,w),  \\
  \Lambda_x^{\mathrm{ml}}(t,w)= \prod_{t_n\in\mathcal{T}^{\mathrm{ow}}[k]} f_{\mathcal{P}}\big(r_x(t_n), \bar{r}_x(t_n)|{t_s[k]=t,w[k]=w}\big),\quad x\in\{A,B\}, \label{Eq:ML_SyncDet_metric}
\end{IEEEeqnarray}
where we exploited the fact that the observations for type-$A$ and type-$B$ molecules and for different sample times are independent. Note that hypotheses $t_s^{\mathrm{ml}}[k]=t$ and $w^{\mathrm{ml}}[k]=w$ affect the signal means $\bar{r}_A(t_n)$ and $\bar{r}_B(t_n)$, cf. (\ref{Eq:ExpMol}). Exploiting again the monotonicity of the logarithm, we can rewrite (\ref{Eq:ML_SyncDet}) as follows 
\begin{IEEEeqnarray}{lll} \label{Eq:ML_log_joint}
[\hat{t}_s^{\mathrm{ml}}[k],\hat{w}^{\mathrm{ml}}[k]] &= \underset{\forall t \in\mathcal{T}[k], w\in\{0,1\} }{\mathrm{argmax}}\,\,&\mathrm{ln}(\Lambda_A^{\mathrm{ml}}(t,w)) + \mathrm{ln}(\Lambda_B^{\mathrm{ml}}(t,w)) \nonumber\\
&=\underset{\forall t \in\mathcal{T}[k], w\in\{0,1\} }{\mathrm{argmax}}&\sum_{t_n\in\mathcal{T}^{\mathrm{ow}}[k]}  
\Big[ r_A(t_n)\mathrm{ln}(\bar{r}_A(t_n))-\bar{r}_A(t_n)-\mathrm{ln}(r_A(t_n)!) \nonumber \\
&& \qquad \quad + r_B(t_n)\mathrm{ln}(\bar{r}_B(t_n))-\bar{r}_B(t_n)-\mathrm{ln}(r_B(t_n)!) \Big].
\end{IEEEeqnarray}
 Although we cannot solve the ML problem in (\ref{Eq:ML_log_joint}) in closed form, we are able to solve it numerically and use it as a performance benchmark for the suboptimal low-complexity~schemes proposed for Framework~2. 
 
The RHS of Fig.~\ref{Fig:Concept} illustrates the synchronization and detection schemes proposed under Framework~2. In particular, the RHSs of Figs.~\ref{Fig:Concept}~a) and b) show the numbers of type-$A$ and type-$B$ molecules observed at the receiver under Framework~2, respectively, for one realization.  The RHS of Fig.~\ref{Fig:Concept} c) illustrates the proposed ML joint synchronization and detection scheme. This figure reveals that for the considered example, the start times of the symbol intervals and the transmitted bits  can be accurately determined by the ML scheme in (\ref{Eq:ML_log_joint}).

\subsubsection{Suboptimal Low-Complexity Schemes}

Analogous to the suboptimal synchronization schemes proposed for Framework~1 in Section~III-B, in the following, we propose three suboptimal joint synchronization and detection schemes for Framework~2, namely an LF scheme, a PO scheme, and a TT scheme.

\begin{figure}
  \centering
\resizebox{1.1\linewidth}{!}{\hspace{-4cm}\psfragfig{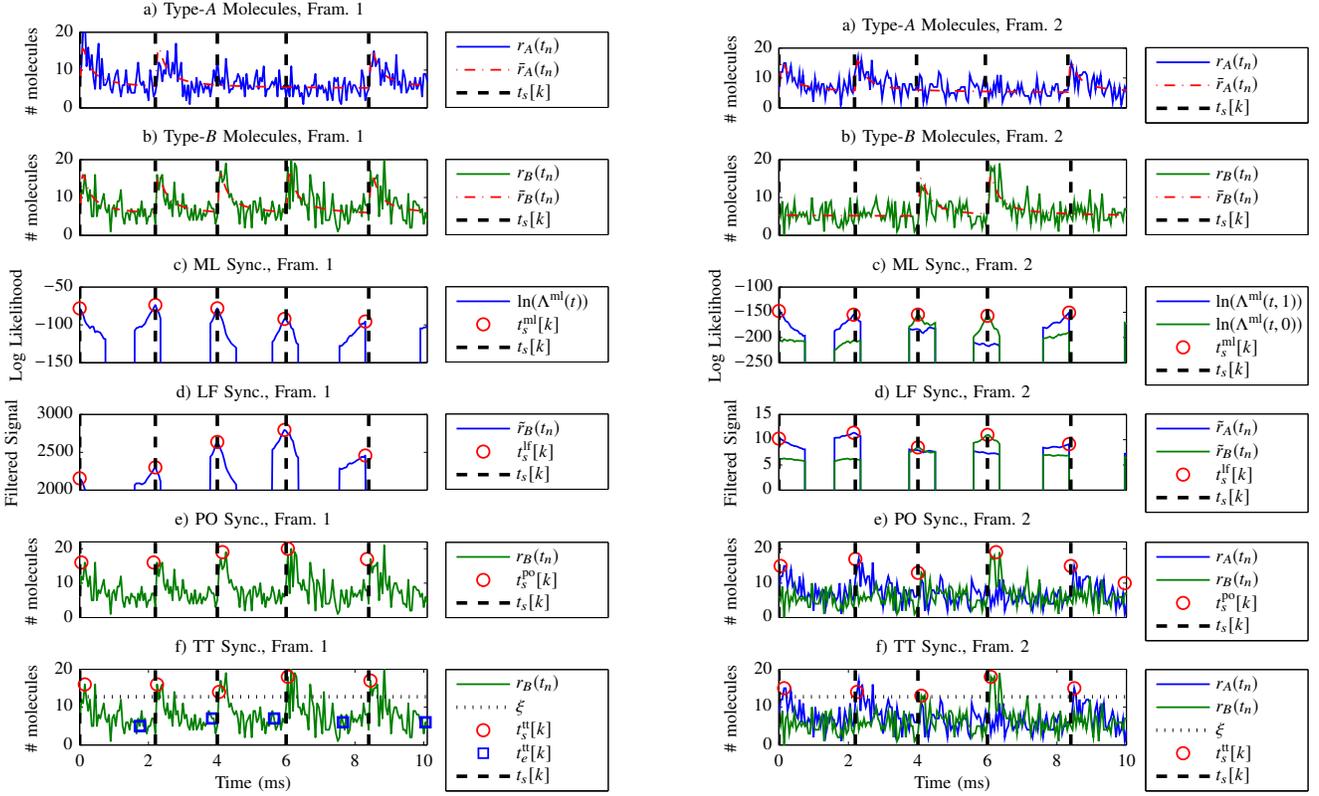}}   \vspace{-1.5cm}
\caption{Illustration of the proposed synchronization schemes for five data symbols $[1,1,0,0,1]$ where the start times of the symbol intervals are $t_s[k]=[0, 2.2, 4, 6, 8.4]$~ms. The details of the adopted simulation setup and the corresponding simulation parameters are given~in~Section~V and Table~III, respectively. }
\label{Fig:Concept}
\end{figure}

\textit{Linear Filter-Based Scheme:}
For this scheme, the received signals for the type-$A$ and type-$B$ molecules are correlated with the corresponding signal means $\bar{r}_A(t_n)$ and $\bar{r}_B(t_n)$, respectively. The filtered signals are denoted by $\tilde{r}_A(t_n)$ and $\tilde{r}_B(t_n)$ and are given by
\begin{IEEEeqnarray}{lll} \label{Eq:FilterSig_joint}
\tilde{r}_x(t) = \frac{1}{c_x^2}\sum_{t_n\in\mathcal{T}^{\mathrm{ow}}[k]} r_x(t_n) \bar{r}_x(t_n),\quad x\in\{A,B\},
\end{IEEEeqnarray}
where $c_x=z_x + {\max}_{t}\,N_xP_x(t),\,\,x\in\{A,B\}$, is a normalization constant. Using the above filtered signals, the proposed LF joint symbol synchronization and detection scheme is given by
\begin{IEEEeqnarray}{lll} \label{Eq:LF_Prot_joint}
\hat{t}_s^{\mathrm{lf}}[k] & = \underset{t_n\in \mathcal{T}[k]}{\mathrm{argmax}} \,\,\max\{\tilde{r}_A(t_n),\tilde{r}_B(t_n)\} 
\IEEEyesnumber\IEEEyessubnumber \\
\hat{w}^{\mathrm{lf}}[k] &= \begin{cases}
1,\quad &\mathrm{if}\,\,\tilde{r}_A(\hat{t}_s^{\mathrm{lf}}[k])\geq \tilde{r}_B(\hat{t}_s^{\mathrm{lf}}[k]) \\
0, & \mathrm{otherwise}.  
\end{cases} \IEEEyessubnumber
\end{IEEEeqnarray}
The LF synchronization and detection scheme in (\ref{Eq:LF_Prot_joint}) is schematically illustrated in the RHS of Fig.~\ref{Fig:Concept} d) for the example scenario. Although the scheme in (\ref{Eq:LF_Prot_joint}) is computationally simpler than the ML scheme in (\ref{Eq:ML_log_joint}), we observe that, for the considered example, the LF scheme can still accurately recover the start times of the symbol intervals and the transmitted bits.

 \textit{Peak Observation-Based  Scheme:}
To further reduce the complexity of joint synchronization and detection under Framework~2, we propose a scheme based on only the peak observation.  Assuming perfect synchronization in the previous symbol intervals, the set of expected time instances where the peak observations for type-$A$ and type-$B$ molecules in symbol interval $k$ can occur is given by 
\begin{IEEEeqnarray}{lll} \label{Eq:Tp_joint}
\mathcal{T}^{\mathrm{p}}[k] = \left[t_s[k-1]+T^{\min}+\min\left\{t_A^{\mathrm{p}},t_B^{\mathrm{p}}\right\},\,
t_s[k-1]+T^{\max}+\max\left\{t_A^{\mathrm{p}},t_B^{\mathrm{p}}\right\} \right],
\end{IEEEeqnarray}
where $t_x^{\mathrm{p}}={\mathrm{argmax}}_{t\geq 0}\,\,P_x(t),\,\,x\in\{A,B\}$.
Hereby, we propose the following PO joint symbol synchronization and detection scheme 
\begin{IEEEeqnarray}{lll} \label{Eq:PeakProt_joint}
\hat{t}_s^{\mathrm{po}}[k] & = \begin{cases}
\Big[\underset{t_n\in \mathcal{T}^{\mathrm{p}}[k]}{\mathrm{argmax}} \,\,r_A(t_n) \Big]-t_A^{\mathrm{p}},\,\,
& \mathrm{if}\,\, \underset{t_n\in \mathcal{T}^{\mathrm{p}}[k]}{\mathrm{max}} \,\,\frac{r_A(t_n)}{c_A}
\geq \underset{t_n\in \mathcal{T}^{\mathrm{p}}[k]}{\mathrm{max}} \,\,\frac{r_B(t_n)}{c_B} \\
\Big[\underset{t_n\in \mathcal{T}^{\mathrm{p}}[k]}{\mathrm{argmax}} \,\,r_B(t_n) \Big]-t_B^{\mathrm{p}},\,\,
& \mathrm{otherwise}
\end{cases}
\IEEEyesnumber\IEEEyessubnumber \\
\hat{w}^{\mathrm{po}}[k] &= \begin{cases}
1,\quad &\mathrm{if}\,\,\frac{r_A(\hat{t}_s^{\mathrm{po}}[k]+t_A^{\mathrm{p}})}{c_A}
\geq \frac{r_B(\hat{t}_s^{\mathrm{po}}[k]+t_B^{\mathrm{p}})}{c_B} \\
0, & \mathrm{otherwise}.  
\end{cases} \IEEEyessubnumber
\end{IEEEeqnarray}
The proposed PO synchronization and detection scheme is shown for the example scenario in  the RHS of Fig.~\ref{Fig:Concept}~e). For the considered example, the PO scheme can accurately  identify the start times of the symbol intervals and the transmitted bits  based on only the two peak observation~samples.

 \textit{Threshold-Trigger Scheme:}
As discussed in Section~III-B, the threshold-trigger mechanism is a common strategy  among living organisms in response to external stimuli  \cite{CellBio}. In Section~III-B, we employed the TT mechanism for determining a detection window, cf. (\ref{Eq:SubOpt_Sync}). In other words, instead of explicitly finding the symbol duration, we employed type-$B$ molecules to specify an interval during which we perform data detection using type-$B$ molecules. Similarly, here, we propose a data detection scheme that does not require the explicit specification of the beginning and the end of the symbol intervals. The main idea behind our proposed TT scheme is that the receiver declares a new symbol interval if either type-$A$ or type-$B$ molecules exceed their respective thresholds. To reduce the probability of false alarm for declaring a new symbol interval, we assume that the receiver enforces a minimum guard time of $T^{\mathrm{dw}}$ before permitting a new symbol interval. On the other hand, we assume that $T^{\mathrm{dw}}\leq T^{\min}$ holds in order to reduce the probability that the receiver misses the start of a new symbol interval.  Taking these considerations into account, the proposed TT scheme is formally given by
\begin{IEEEeqnarray}{lll} \label{Eq:TT_joint}
\hat{t}_s^{\mathrm{tt}}[k]&=\underset{t_n>\hat{t}_s^{\mathrm{tt}}[k-1]+T^{\mathrm{dw}}}{\mathrm{min}}\,\,
t_n \big | \max\left\{\frac{r_A(t_n)}{c_A},\frac{r_B(t_n)}{c_B}\right\} \geq \xi \IEEEyesnumber\IEEEyessubnumber \\
\hat{w}^{\mathrm{tt}}[k] &= \begin{cases}
1,\quad &\mathrm{if}\,\,\frac{r_A(\hat{t}_s^{\mathrm{tt}}[k])}{c_A} \geq \frac{r_B(\hat{t}_s^{\mathrm{tt}}[k])}{c_B} \\
0, & \mathrm{otherwise}. 
\end{cases} \IEEEyessubnumber
\end{IEEEeqnarray}
In the RHS of Fig.~\ref{Fig:Concept} f), we illustrate the proposed TT synchronization and detection scheme for the example scenario. For the considered example, the TT scheme is able to accurately  identify the beginning of the symbol intervals and the transmitted bits.

%

\section{Comparison of Proposed Synchronization Schemes and Consequences of Synchronization Errors}

In this section, we first compare the characteristics of the proposed synchronization schemes in detail. Subsequently, we discuss deletion and insertion errors caused by imperfect synchronization.

\subsection{Comparison of the Proposed Synchronization Schemes}
In the following, we compare several different aspects of the proposed synchronization schemes.

\subsubsection{Required A Priori Knowledge, Constraints, and Relative Complexity}
Here, we focus on the comparison of the proposed symbol synchronization schemes, i.e., the estimation of $t_s[k]$, and we do not consider the detection stage, i.e., the detection of $w[k]$. Table~I summarizes the required a priori knowledge, the underlying constraints, and the complexity of the proposed symbol synchronization schemes. For the considered MC system, the channel is characterized by $P_x(t)$  and $z_x$, $x\in\{A,B\}$.  The ML synchronization scheme requires full knowledge of the channel characteristics of the molecules used for synchronization. In other words, the ML synchronization scheme for Framework~1 requires knowledge of $P_B(t)$  and $z_B$ whereas the ML synchronization scheme for Framework~2 requires knowledge of $P_A(t)$, $P_B(t)$, $z_A$,  and $z_B$. For both frameworks, full knowledge of the MC channel is also needed  for the proposed LF estimators; however, the complexity associated with filtering in (\ref{Eq:FilterProt_metric}) and (\ref{Eq:FilterSig_joint}), i.e., linear operations, is lower than that required for computing the ML metrics in  (\ref{Eq:ML_Sync_metric})  and (\ref{Eq:ML_SyncDet_metric}).  Compared to the ML and LF schemes, the proposed PO and TT synchronization schemes require less a priori information about the channel and entail a lower computational complexity.  We note that the parameters required for all proposed schemes are constant for the coherence time of the MC channel. Hence, the receiver can obtain them offline at the beginning of transmission and use them for online symbol synchronization as long as the MC channel statistics remain unchanged.  We further note  that unlike for the ML and LF synchronization schemes, for which the strict constraint $T^{\max}\leq 2T^{\min}$ has to hold, the proposed PO and TT synchronization schemes do not have this~restriction. 

\begin{table}
\label{Table:Knowledge}
\caption{Requirements for the applicability of the Proposed Symbol Synchronization Schemes.} 
\begin{center}
\scalebox{0.7} { 
\begin{tabular}{|| c | c | c| c ||}
  \hline
 Sync. Scheme &  A Priori Knowledge & Constraints & Relative Complexity \\ \hline \hline 
 \multicolumn{4}{||c||}{Framework~1} \\ \hline  
 ML Scheme & $P_B(t)$ and $z_B$   &   $T^{\max}\leq 2T^{\min}$  & High  \\ \hline
 LF Scheme & $P_B(t)$ and $z_B$    &   $T^{\max}\leq 2T^{\min}$ & Moderate  \\ \hline
 PO Scheme & $t^{\mathrm{p}}$  & $-$ & Low  \\ \hline
 TT Scheme & $\xi$  &  $-$  & Low  \\ \hline \hline 
 \multicolumn{4}{||c||}{Framework~2} \\ \hline  
 ML Scheme & $P_A(t)$,  $P_B(t)$, $z_A$, and $z_B$   &   $T^{\max}\leq 2T^{\min}$  & High  \\ \hline
 LF Scheme & $P_A(t)$,  $P_B(t)$, $z_A$, and $z_B$   &   $T^{\max}\leq 2T^{\min}$  & Moderate  \\ \hline
 PO Scheme & $t^{\mathrm{p}}$, $c_A$, and $c_B$  & $-$ & Low  \\ \hline
 TT Scheme & $\xi$, $c_A$, and $c_B$    &  $-$  & Low  \\ \hline  
\end{tabular}
}
\end{center}
\vspace{-0.3cm}
\end{table}

\subsubsection{Average Molecule Consumption} 
In this paper, we assume that the maximum number of molecules that the transmitter can release in each symbol interval is fixed, i.e., a per-symbol power constraint is applied. Nevertheless, the average number of molecules employed under Framework~1 is larger than that under Framework~2. In particular, denoting the average number of molecules released per symbol interval by $\bar{N}$ and assuming that the binary symbols are equiprobable, we obtain $\bar{N}=\frac{1}{2}N_A+N_B$ and $\bar{N}=\frac{1}{2}N_A+\frac{1}{2}N_B$ for Framework~1 and Framework~2, respectively. It is also possible to impose a constraint on the maximum number of  molecules released per symbol interval, $N_A$ and $N_B$, and a constraint on the average number of overall released molecules, $\bar{N}$, of course. To formulate these constraints more rigorously, let us assume that the transmitter releases on average $\beta\bar{N}$ type-$A$ molecules and $(1-\beta)\bar{N}$ type-$B$ molecules per symbol interval, i.e., the overall average number of molecules released per symbol interval is constrained by $\bar{N}$. Thereby, the maximum number of  molecules released per symbol interval is obtained as $N_A=2\beta \bar{N}$ and $N_B=(1-\beta) \bar{N}$ for Framework~1 and $N_A=2\beta \bar{N}$ and $N_B=2(1-\beta) \bar{N}$ for Framework~2. In the simulation results, we consider this scenario in Fig.~\ref{Fig:BER_Beta} and show that $\beta$ can be optimized for performance maximization. 

\subsubsection{Extension to Multi-Node Synchronization}
In this paper, we consider a point-to-point MC system. We note that an advantage of Framework~1 over Framework~2 is that the  synchronization schemes proposed for Framework~1 are also applicable  to the broadcast channel, i.e., when one transmitter wishes to communicate with multiple receivers. In this case, the transmitter may employ different types of information molecules for each receiver, e.g., type $A_1,A_2,\dots$, and $A_M$ molecules for receivers $1,2,\dots$, and $M$, respectively. However, in such a broadcast channel, for Framework~1, only one type of synchronization molecule, e.g., type $B$, is sufficient for synchronization of all links, provided that the transmitter employs the same symbol interval for all types of emitted molecules. Therefore, each receiver can independently apply the synchronization and detection schemes proposed  for Framework~1. Hence, an important benefit of Framework~1 is that  as the number of receivers increases, the total synchronization overhead (in terms of the  resources required for synchronization) remains constant. 

\subsection{Insertion and Deletion Errors}

\begin{figure}
  \centering
 \scalebox{0.8}{
\pstool[width=1\linewidth]{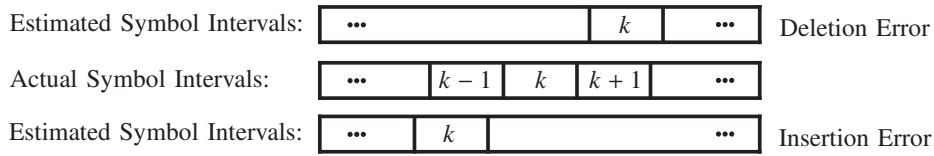}{
\psfrag{k}[c][c][1]{$k$}
\psfrag{k1}[c][c][1]{$k-1$}
\psfrag{k2}[c][c][1]{$k+1$}
\psfrag{A}[l][c][1]{Actual Symbol Intervals:}
\psfrag{D}[l][c][1]{Estimated Symbol Intervals:}
\psfrag{B}[l][c][1]{Insertion Error}
\psfrag{C}[l][c][1]{Deletion Error}
} } \vspace{-0.3cm} 
\caption{Illustration of deletion and insertion errors for the $k$-th symbol interval arising from imperfect symbol synchronization  assuming no deletion or insertion has occurred for symbol intervals $k'<k,\,\forall k'$, i.e., $k^{\mathrm{offset}}=0$. \vspace{-0.2cm}}
\label{Fig:InsDelError}
\end{figure}

A common challenge of imperfect symbol synchronization are deletion and insertion errors \cite{InsDelMarkerCode,InsDelCode}. A deletion error occurs if the adopted synchronization protocol fails to identify the start of a symbol interval, and an insertion error occurs if a false alarm introduces an additional symbol interval. We note that deletion and insertion errors may have a severe impact on the BER performance. For instance, a single insertion or deletion error shifts the positions of all subsequent symbols and significantly deteriorates the BER calculated based on a symbol-by-symbol comparison of the transmitted and detected data. Insertion and deletion errors are schematically illustrated in Fig.~\ref{Fig:InsDelError} and mathematically defined as
\begin{IEEEeqnarray}{lll} \label{Eq:DelInsError}
\begin{cases}
\hat{t}_s[k] \geq t_s[k+k^{\mathrm{offset}}+1],\quad &\text{deletion error for symbol $k$} \\
\hat{t}_s[k+1] \leq t_s[k+k^{\mathrm{offset}}],\quad &\text{insertion error for symbol $k$} 
\end{cases}
\end{IEEEeqnarray}
where $\hat{t}_s[k]$ is an estimate for $t_s[k]$ for one of the proposed synchronization schemes and $k^{\mathrm{offset}}$ is the number of deletion events minus the number of insertion events that have occurred for $\forall k'<k$. To cope with this challenge in conventional communication systems, special codes were designed which are capable of  correcting codewords corrupted by insertions and deletions \cite{InsDelMarkerCode,InsDelCode}. For instance, in \cite{InsDelMarkerCode}, F. Sellers proposed to periodically insert a synchronizing marker sequence at the beginning/end of each data block. By searching for the markers in their expected positions, the decoder can detect and subsequently correct single insertions or deletions between successive markers. In a similar manner, multiple synchronization errors can be corrected by using longer markers, at the expense of additional redundancy \cite{InsDelCode}. 

In this paper, to illustrate the impact and mitigation of insertion and deletion errors in MC systems, we adopt the codes proposed in \cite{InsDelMarkerCode}. In particular, let $\mathbf{W}$ be the original data sequence and $\mathbf{M}$ be a marker of length $l$ which is inserted periodically at the end of each $L$ data bits to create the encoded data symbols $\bar{\mathbf{W}}$. For instance, if the data sequence is $\mathbf{W} = 101010\,\,001011$, $L=6$, and $\mathbf{M}=100$, the encoded data symbols will be $\bar{\mathbf{W}} = 101010\textbf{100}\,\,001011\textbf{100}$. The decoder for this example code is given in Table~II. It was shown in \cite{InsDelMarkerCode} that this code can correct one deletion or one insertion error. In Section~V, we study the effectiveness of the aforementioned simple deletion/insertion error correction code in MC systems employing the proposed synchronization schemes. Designing  deletion/insertion codes specifically for MC systems is an interesting topic for future research.

\begin{table}
\label{Table:InsDel}
\caption{Receiver Decision for a Deletion/Insertion Code with Marker $\mathbf{M}=100$ \cite{InsDelMarkerCode}.} 
\begin{center}
\scalebox{0.7} { 
\begin{tabular}{|| c | c | c ||}
  \hline
 Detected Marker &  Decision & Receiver Action \\ \hline \hline 
 $100$ & No error &  No action  \\ \hline
  $000$/$001$ & Deletion error in data & Add one bit before the marker  \\ \hline
 $010$/$110$ & Insertion error in data & Remove one bit before the marker  \\ \hline
  $101$ & Deletion/insertion error in marker & No action  \\ \hline
  $111$/$011$ & Deletion/insertion error in marker & No action  \\ \hline
\end{tabular}
}
\end{center}
\vspace{-0.3cm}
\end{table}

\section{Simulation Results}

In the following, we first present the MC setup used for generating the simulation results. Next, we evaluate the performance of the proposed schemes in terms of the synchronization estimation error and the end-to-end BER.

\subsection{Simulation Setup and Performance Metrics}
For simplicity, we assume instantaneous molecule release by a point source transmitter, consider an unbounded three-dimensional environment, and employ the reactive receiver model recently developed in \cite{Arman_ReactReciever} for the calculation of $P_x(t),\,\,x\in\{A,B\}$. Moreover, we assume that $t_s[k]$ is uniformly distributed in $\mathcal{T}[k]$, i.e.,  the length of each symbol interval is an RV uniformly distributed in the interval $[T^{\min},T^{\max}]$. Let $\bar{T}^{\mathrm{symb}}$ be the average symbol duration, i.e., $\bar{T}^{\mathrm{symb}}=\frac{T^{\max}+T^{\min}}{2}$. To be able to easily control the variability of the symbol intervals with a single parameter, we assume that $T^{\min}=(1-\alpha)\bar{T}^{\mathrm{symb}}$ and $T^{\max}=(1+\alpha)\bar{T}^{\mathrm{symb}}$ hold, where $\alpha\in[0,1]$ in general\footnote{In order to enforce the constraint $T^{\max}\leq 2T^{\min}$ for the proposed ML and LF schemes, $\alpha\leq\frac{1}{3}$ has to hold.}. Furthermore, we consider blocks of $K=20$ symbol intervals and  average our results over $10^6$ blocks.  Unless stated otherwise, we adopt the default values of the system parameters given in Table~III. Moreover, we assume $z_x=5,\,\,x\in\{A,B\}$, and change the number of molecules released by the transmitter to obtain different SNRs. For instance, for the default  system parameters in Table~III, we obtain $\text{SNR}_A=\text{SNR}_B=3$ dB.

\begin{table}
\label{Table:Parameter}
\caption{Default Values for Simulation Parameters \cite{Arman_ReactReciever}.\vspace{-0.2cm}} 
\begin{center}
\scalebox{0.6} { 
\begin{tabular}{|| c | c || c | c ||}
  \hline
 Parameter &  Definition &  Value \\ \hline \hline 
$N_A,N_B$ & Number of released type-$A$ and -$B$   molecules  & $10^3$ molecules  \\  \hline
$z_A,z_B$ & Expected number of interfering type-$A$ and -$B$ molecules at the receiver & $5$ molecules  \\  \hline
$n_A,n_B$ & Number of type-$A$ and -$B$ receptors & $10^3$ receptors   \\  \hline
$D$ & Diffusion coefficient of type-$A$ and -$B$ molecules & $5\times 10^{-9}$ $\text{m}^2\cdot\text{s}^{-1}$   \\  \hline
$r_0$ & Distance between transmitter and receiver & $2$ $\mu$m \\  \hline
$r_r$ & Radius of the spherical receiver & $1$ $\mu$m   \\  \hline
$k_f$ & Forward reaction rate for molecule binding& $50\times 10^{-14}$  $\text{m}^3\cdot\text{moleclue}^{-1}\cdot\text{s}^{-1}$  \\  \hline
$k_r$ & Backward reaction rate for molecule binding & $10\times 10^{4}$ $\text{s}^{-1}$   \\  \hline   
$\bar{T}^{\mathrm{symb}}$ & Average length of symbol interval & $2$ ms   \\  \hline 
$\alpha$ & Variability parameter for the length of the symbol intervals & $0.2$   \\  \hline          
$\Delta t$ & Interval between two samples & $50$ $\mu$s   \\  \hline   
$T^{\mathrm{dw}}$ & Length of the detection window for TT synchronization & $T^{\mathrm{min}}$  \\  \hline     
\end{tabular}
}
\end{center}
\vspace{-0.3cm}
\end{table}

In order to compare the performances of the considered synchronization schemes, we define the normalized synchronization error as  
\begin{IEEEeqnarray}{lll} 
\bar{e}_t[k] = \frac{\hat{t}_s[k]-t_s[k]}{\bar{T}^{\mathrm{symb}}}.
\end{IEEEeqnarray}
 Moreover, the end-to-end BER in one block is computed as
\begin{IEEEeqnarray}{lll} 
\text{BER} = \frac{1}{K}\sum_{k=1}^K \left|\hat{w}[k]-w[k]\right|,
\end{IEEEeqnarray}
where $|\cdot|$ denotes the absolute value of a number. Finally, we note that for the proposed TT schemes, the value of threshold $\xi$ is optimized for optimal performance, i.e., minimum BER in Figs.~\ref{Fig:BER_SNR}, \ref{Fig:BER_Beta}, and \ref{Fig:BER_SNR_InsDel} and minimum $\mathcal{E}\{|\bar{e}_t[k]|\}$ for the remaining figures.

\subsection{Synchronization Error}

\begin{figure}[t]
  \centering  
    \resizebox{1.08\linewidth}{!}{\hspace{-2.4cm}
\psfragfig{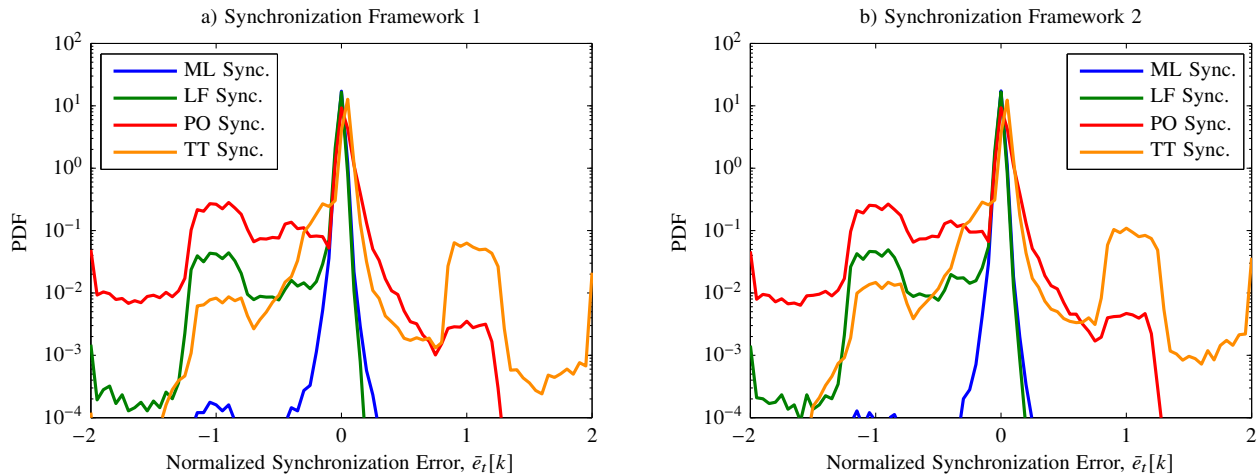}}
\caption{ Estimated PDF (histogram) of the normalized synchronization error, $\bar{e}_t[k]$,  for $\alpha=0.2$, $\bar{T}^{\mathrm{symb}}=2$ ms, and $\text{SNR}_A=\text{SNR}_B = 3$ dB. A bin size of $0.05$ is used for computing the histograms.}
\label{Fig:PDF}  
\end{figure}

\begin{figure}[t]
  \centering
    \resizebox{1.08\linewidth}{!}{\hspace{-2.7cm}
\psfragfig{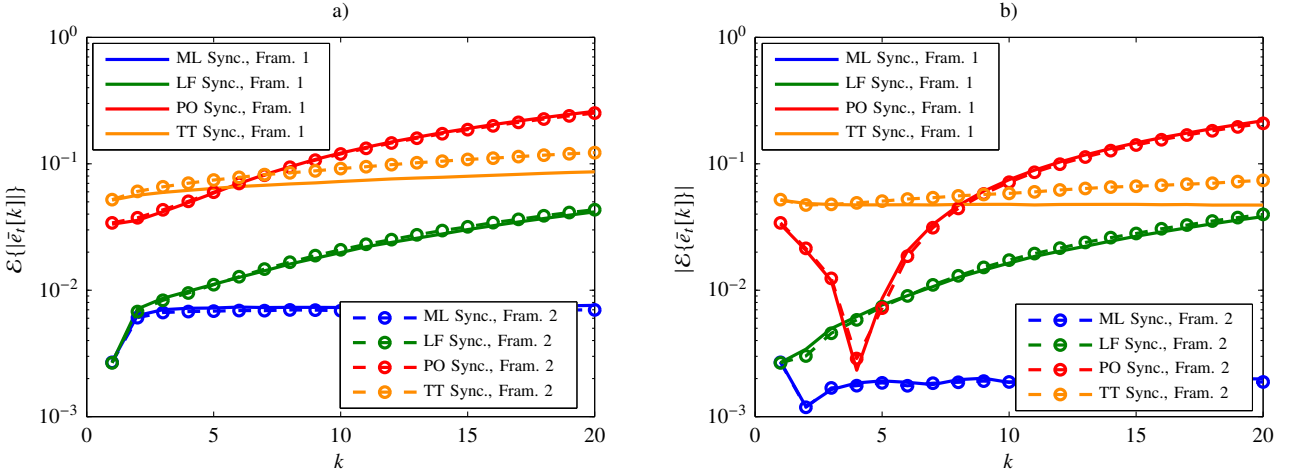}}
\caption{ a) Average absolute normalized synchronization error, $\mathcal{E}\{|\bar{e}_t[k]|\}$,  and b) absolute average normalized synchronization error, $|\mathcal{E}\{\bar{e}_t[k]\}|$, versus symbol index $k$ for $\alpha=0.2$, $\bar{T}^{\mathrm{symb}}=2$ ms, and $\text{SNR}_A=\text{SNR}_B = 3$ dB. }
\label{Fig:Eerror}  \vspace{-0.5cm}
\end{figure}

\begin{figure}[t]
  \centering
    \resizebox{1.08\linewidth}{!}{\hspace{-2.7cm}
\psfragfig{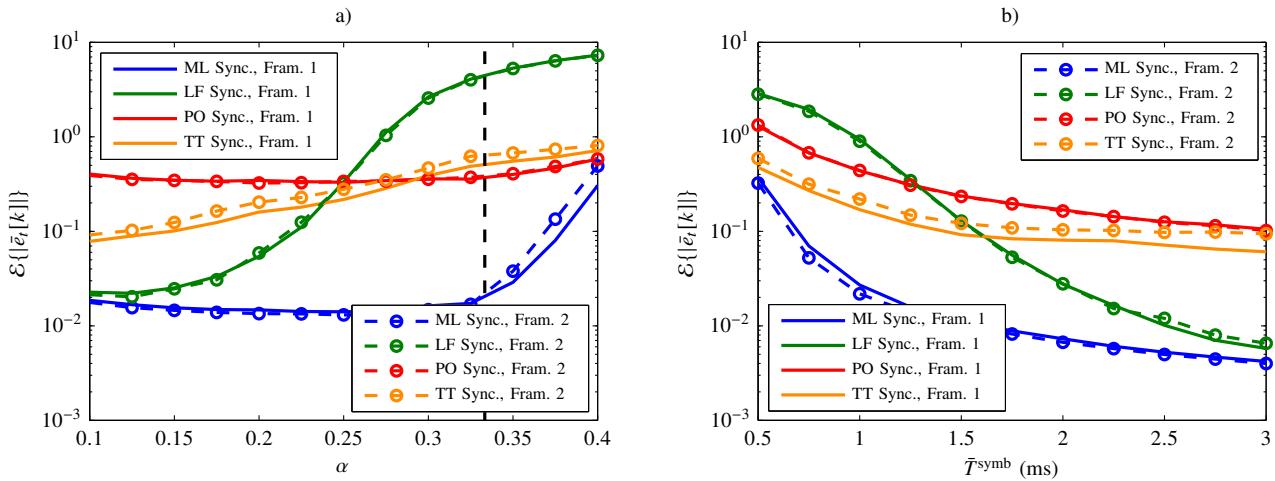}}
\caption{ a) Average absolute normalized synchronization error, $\mathcal{E}\{|\bar{e}_t[k]|\}$, versus parameter $\alpha$ for $\bar{T}^{\mathrm{symb}}=2$ ms and $\text{SNR}_A=\text{SNR}_B = 3$ dB, and b) average absolute normalized synchronization error, $\mathcal{E}\{|\bar{e}_t[k]|\}$, versus symbol duration $\bar{T}^{\mathrm{symb}}$ for $\alpha=0.2$ and $\text{SNR}_A=\text{SNR}_B = 3$ dB. The vertical line in a) corresponds to $\alpha=\frac{1}{3}$ beyond which the constraint $T^{\max}\leq 2T^{\min}$ required for the proposed ML and LF schemes does not hold.}
\label{Fig:Error_ISI}  \vspace{-0.5cm}
\end{figure}

 In Fig.~\ref{Fig:PDF}, we show the histogram of $\bar{e}_t[k]$ for $\alpha=0.2$, $\bar{T}^{\mathrm{symb}}=2$~ms, and $\text{SNR}_A=\text{SNR}_B = 3$~dB. In the following, we highlight some interesting observations from this figure.  
 \begin{itemize}
 \item First, for both frameworks, we observe that the peaks of the probability density function (PDF) for the ML, LF, and PO synchronization schemes are centered at $\bar{e}_t[k]=0$ whereas for the TT synchronization scheme, the peak of the PDF occurs at a positive value of $\bar{e}_t[k]$. This is expected since the TT synchronization scheme does not aim to estimate the start of the symbol intervals and only determines when the number of  molecules bound to the receptors is above threshold $\xi$.
 \item Fig.~\ref{Fig:PDF}  also reveals the presence of insertion and deletion errors for the proposed synchronization schemes for both frameworks, cf. Subsection~IV-A. In particular,  small values of $|\bar{e}_t[k]|$ mean that there are no deletion and no insertion errors, whereas  large and  small values of $\bar{e}_t[k]$ (i.e., $\bar{e}_t[k]>0.5$ and $\bar{e}_t[k]<-0.5$) correspond to deletion and insertion errors,  respectively. Fig.~\ref{Fig:PDF} shows that the probabilities of insertion and deletion errors are not equal for the proposed synchronization schemes. Moreover, we see from Fig.~\ref{Fig:PDF} that for both frameworks, deletion errors are more likely to occur for the TT synchronization scheme than for the PO synchronization scheme since the probability that large values of $\bar{e}_t[k]$ occur is higher for the PO scheme than for the  TT scheme. On the other hand, we observe that error events $\bar{e}_t[k]>0.5$ are unlikely for the ML and LF synchronization schemes which suggests that  deletion errors do not occur for these schemes. In contrast, by considering the range $\bar{e}_t[k]<-0.5$ in Fig.~\ref{Fig:PDF}, we note that the insertion error probabilities increase from ML to TT to LF to PO synchronization for both frameworks. 
 \item We note that a direct comparison of the two frameworks based on Fig.~\ref{Fig:PDF} is not straightforward. Nevertheless, by visually comparing the curves in Figs.~\ref{Fig:PDF} a) and b), we can observe that the synchronization error PDFs for the respective synchronization schemes for Framework~1 are similar to those for Framework~2. As will be verified in Figs.~\ref{Fig:Eerror} and \ref{Fig:Error_ISI}, in terms of synchronization error, Framework~1 indeed achieves a similar performance as  Framework~2. 
\end{itemize}    
 
In order to compare the proposed synchronization schemes more quantitatively, in Fig.~\ref{Fig:Eerror}  a), we plot the mean absolute error (MAE), $\mathcal{E}\{|\bar{e}_t[k]|\}$, versus symbol index $k$ for $\alpha=0.2$, $\bar{T}^{\mathrm{symb}}=2$ ms, and $\text{SNR}_A=\text{SNR}_B = 3$~dB. We observe from Fig.~\ref{Fig:Eerror}  a) that the MAE increases with increasing symbol index for the proposed suboptimal schemes which is due to the accumulation of errors. In contrast, for the ML scheme, the MAE remains almost constant in successive  symbol intervals which indicates that even if an error occurs in one symbol interval, the ML scheme is able to reasonably recover synchronization for the next symbols without a noticeable performance degradation. Moreover, except for the TT scheme which achieves a better performance under Framework~1 than under Framework~2, the MAE performance of the proposed synchronization schemes is almost identical for both frameworks.

In order to study the bias of the proposed estimators, in Fig.~\ref{Fig:Eerror}  b), we show the absolute mean error, $|\mathcal{E}\{\bar{e}_t[k]\}|$, versus symbol index $k$ for $\alpha=0.2$, $\bar{T}^{\mathrm{symb}}=2$ ms, and $\text{SNR}_A=\text{SNR}_B = 3$~dB.  From this figure, we conclude that all proposed estimators are biased. We note that for the ML and LF schemes, $\mathcal{E}\{\bar{e}_t[k]\}$ is positive for $k=1$ and negative for $k\geq 2$; for  the PO schemes, $\mathcal{E}\{\bar{e}_t[k]\}$ is positive for $k\leq 4$ and negative for $k\geq 5$; and for the TT scheme, $\mathcal{E}\{\bar{e}_t[k]\}$ is positive for all $k$. Overall, for large $k$, the mean of the synchronization error, $\mathcal{E}\{\bar{e}_t[k]\}$, is negative for all schemes except the TT scheme and $|\mathcal{E}\{\bar{e}_t[k]\}|$ increases with symbol index $k$ particularly for the suboptimal LF and PO schemes.

Next, we investigate the effect of the variability of the symbol duration on the synchronization error performance. In particular, in Fig.~\ref{Fig:Error_ISI} a), we plot the MAE, $\mathcal{E}\{|\bar{e}_t[k]|\}$, versus parameter $\alpha$ for $\bar{T}^{\mathrm{symb}}=2$ ms and $\text{SNR}_A=\text{SNR}_B = 3$ dB. We observe that for the ML schemes, the MAE remains almost constant for $\alpha\leq \frac{1}{3}$ and considerably deteriorates for $\alpha > \frac{1}{3}$. The reason for this is that for $\alpha > \frac{1}{3}$, the constraint $T^{\max}\leq 2T^{\min}$, which is required for the proposed ML schemes, does not hold. Another interesting observation from Fig.~\ref{Fig:Error_ISI}~a) is that the LF scheme outperforms the less complex PO and TT schemes only if  $\alpha < 0.25$ holds.

In order to study the effect of ISI on the error performance, in Fig.~\ref{Fig:Error_ISI} b), we show MAE, $\mathcal{E}\{|\bar{e}_t[k]|\}$, versus symbol duration $\bar{T}^{\mathrm{symb}}$ for $\alpha=0.2$ and $\text{SNR}_A=\text{SNR}_B = 3$ dB. As expected, the error performance of all proposed schemes improves with increasing symbol duration  since the ISI decreases as $\bar{T}^{\mathrm{symb}}$ increases. Moreover, the performance of the LF schemes is very close to that of the ML schemes when the ISI is negligible. On the other hand, if the ISI is severe, the LF schemes may  be outperformed even by the simple PO and TT schemes. The higher sensitivity of the LF schemes to small $\bar{T}^{\mathrm{symb}}$  compared to the  PO and TT schemes can be attributed to the fact that the filtering operation increases the length of the overall impulse response, and hence introduces additional ISI. Therefore, a decrease of $\bar{T}^{\mathrm{symb}}$ deteriorates the performance of the LF scheme more severely than  that of the  PO and TT schemes.

\subsection{BER Performance}

\begin{figure}[t]
  \centering
    \resizebox{1.08\linewidth}{!}{\hspace{-2.7cm}
\psfragfig{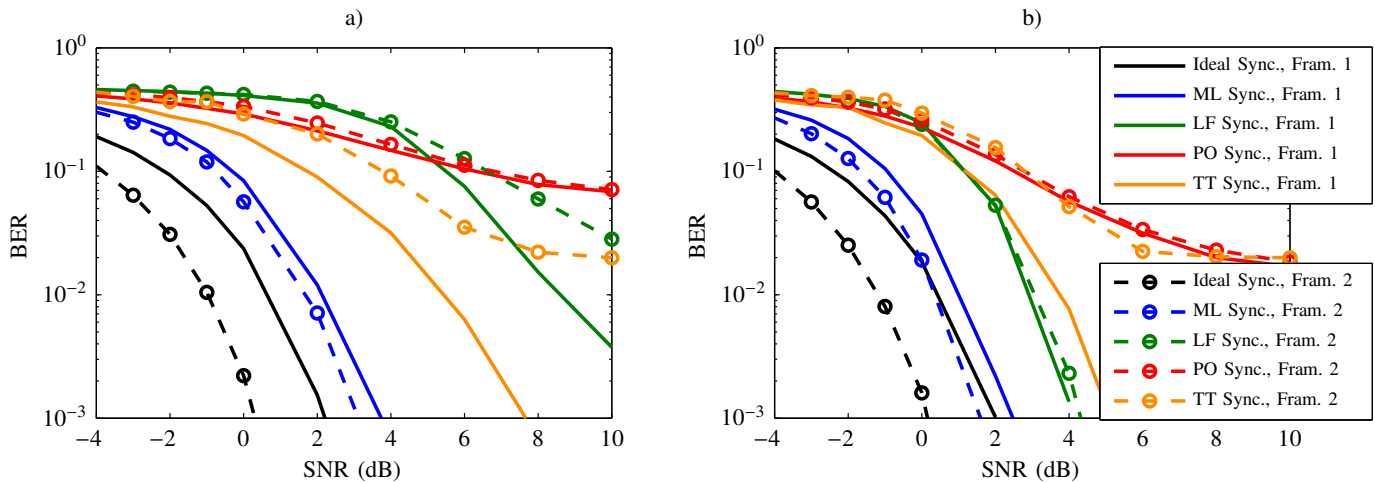}}
\caption{ BER versus $\text{SNR}=\text{SNR}_A=\text{SNR}_B$  for $\alpha=0.2$ and a) $\bar{T}^{\mathrm{symb}}=1$ ms (strong ISI), b) $\bar{T}^{\mathrm{symb}}=2$ ms (weak ISI). }
\label{Fig:BER_SNR}  \vspace{-0.5cm}
\end{figure}

Next, we study the performance of the proposed synchronization schemes in terms of the end-to-end BER. As performance upper bound, we consider the case of perfect symbol synchronization where the beginning of the symbol intervals is perfectly known at the receiver and the ML detection in (\ref{Eq:ML_Det_Log}) and (\ref{Eq:ML_log_joint}) is adopted for Frameworks~1 and 2, respectively. In Fig.~\ref{Fig:BER_SNR}, the BER is shown versus $\text{SNR}=\text{SNR}_A=\text{SNR}_B$ for $\alpha=0.2$ and a) $\bar{T}^{\mathrm{symb}}=1$ ms (strong ISI), b) $\bar{T}^{\mathrm{symb}}=2$ ms (weak ISI). We highlight the following observations from Fig.~\ref{Fig:BER_SNR}.
\begin{itemize}
\item From Fig.~\ref{Fig:BER_SNR}~a), we observe that there is an SNR gap of approximately $2$ dB between the BERs of the ML synchronization schemes and the upper bound which is solely due to imperfect synchronization. This gap becomes smaller in Fig.~\ref{Fig:BER_SNR}~b), particularly for Framework~1, as there is less ISI  and hence the quality of synchronization improves. The BER of the LF scheme in Fig.~\ref{Fig:BER_SNR}~b) is significantly improved compared  to that in Fig.~\ref{Fig:BER_SNR}~a) due to the weaker ISI. This trend was also expected from Fig.~\ref{Fig:Error_ISI}~b) based on the synchronization errors for  $\bar{T}^{\mathrm{symb}}=1$ and $\bar{T}^{\mathrm{symb}}=2$ ms.
\item  As expected, joint ML synchronization and detection in (\ref{Eq:ML_log_joint}) under Framework~2 outperforms independent ML synchronization in (\ref{Eq:ML_log}) and ML detection in (\ref{Eq:ML_Det_Log}) under Framework~1. However, the suboptimal schemes for Framework~2 are all outperformed by the respective suboptimal schemes for Framework~1. This is due to the fact that for the suboptimal synchronization schemes for Framework~1, we employ optimal ML detection in (\ref{Eq:ML_Det_Log}) whereas for the suboptimal schemes for Framework~2,  synchronization and detection are performed jointly and are strictly suboptimal.
\item Figs.~\ref{Fig:BER_SNR}~a) and b) reveal that  the TT scheme for Framework~1 offers a good performance for both ISI scenarios. Therefore, the TT scheme for Framework~1 might be a good option for practical MC systems with limited computational capabilities, since it provides a favorable tradeoff between complexity and performance.
\end{itemize}

\begin{figure}[t]
  \centering
    \resizebox{1.08\linewidth}{!}{\hspace{-2.7cm}
\psfragfig{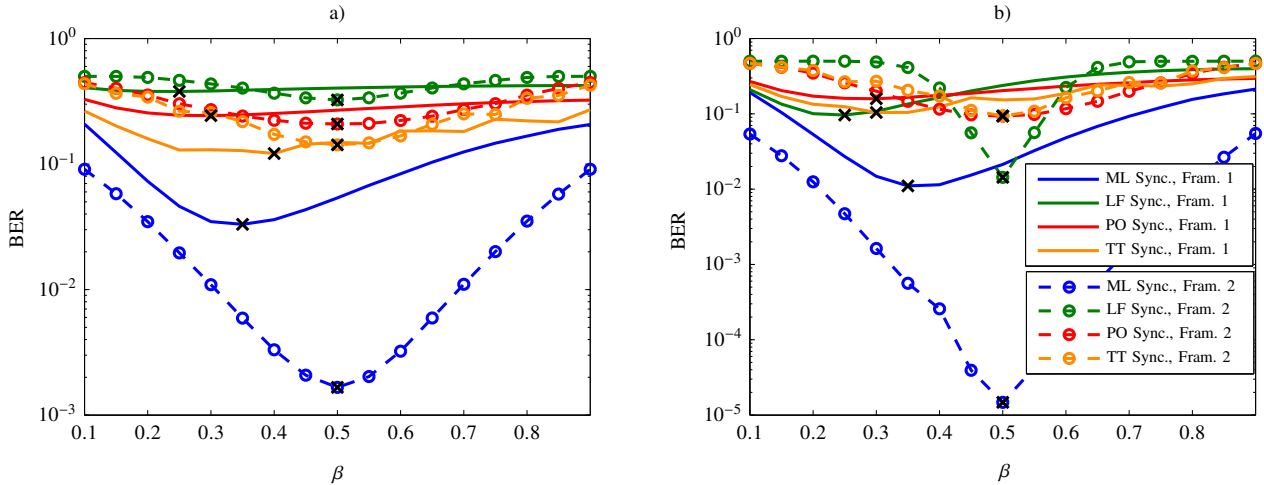}}
\caption{ BER versus parameter $\beta$ for $\alpha=0.2$, $\bar{N}=10^3$ per symbol, and a) $\bar{T}^{\mathrm{symb}}=1$ ms (strong ISI), b) $\bar{T}^{\mathrm{symb}}=2$ ms (weak ISI). The cross markers denote the minimum BER value for each curve. }
\label{Fig:BER_Beta}  \vspace{-0.5cm}
\end{figure}

Recall that in the system model, we assume that the maximum number of type-$A$ and type-$B$ molecules that the transmitter can release in a given symbol interval is $N_A$ and $N_B$, respectively. Moreover, following the discussion in Section~IV-B-2, one can also constrain the average number of molecules that the transmitter releases by $\bar{N}$. Here, we consider both constraints and aim to optimize the fractions of type-$A$ molecules, i.e., $\beta$, and type-$B$ molecules, i.e., $1-\beta$, see  Section~IV-B-2 for details. In Fig.~\ref{Fig:BER_Beta}, we show the BER versus parameter $\beta$ for $\alpha=0.2$, $N^{\max}=10^3$ per symbol, and a) $\bar{T}^{\mathrm{symb}}=1$ ms (strong ISI) and b) $\bar{T}^{\mathrm{symb}}=2$ ms (weak ISI). As expected, the optimal $\beta$ for all schemes in Framework~2 is $0.5$ which is due to the symmetry of the system with respect to type-$A$ and type-$B$ molecules. Interestingly, for all schemes in Framework~1, the optimal value of $\beta$ is strictly lower than $0.5$. This suggests that for the overall BER performance, it is beneficial to allocate more molecules to symbol synchronization than to data detection. Moreover, the BER performance is considerably better for the case of weak ISI (Fig.~\ref{Fig:BER_Beta} b)) compared to that of strong ISI (Fig.~\ref{Fig:BER_Beta} a)).

\subsection{Insertion and Deletion Errors}

\begin{figure}[t]
  \centering
    \resizebox{1.08\linewidth}{!}{\hspace{-2.7cm}
\psfragfig{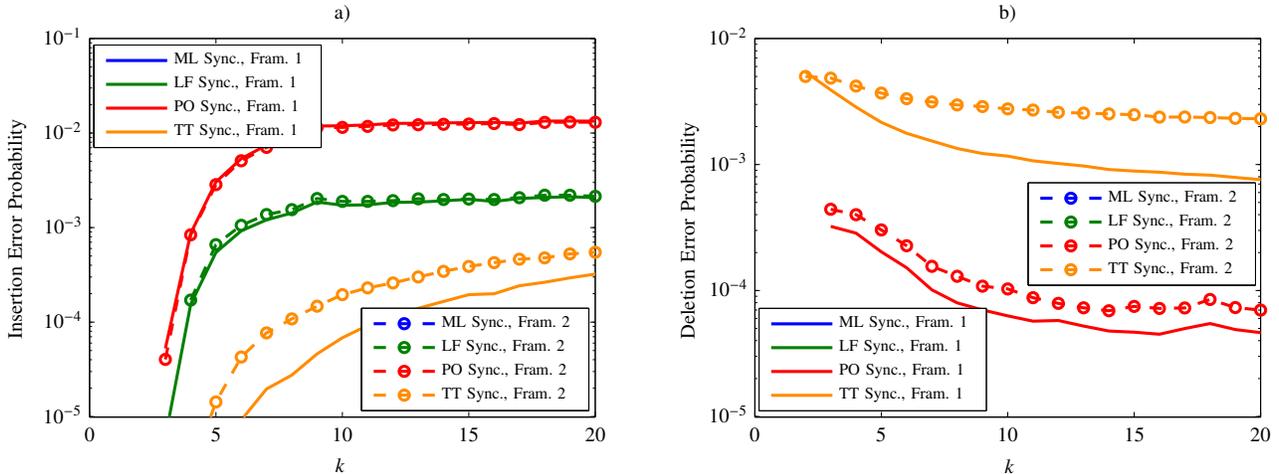}}
\caption{ a) Insertion error probability and b) deletion error probability versus symbol index $k$ for $\alpha=0.2$, $\bar{T}^{\mathrm{symb}}=2$ ms, and $\text{SNR}_A=\text{SNR}_B = 3$ dB. }
\label{Fig:InsDel}  \vspace{-0.5cm}
\end{figure}

In this subsection, we study the deletion and insertion errors of the proposed synchronization schemes as defined in (\ref{Eq:DelInsError}). In particular, in Figs.~\ref{Fig:InsDel} a) and b), we show the insertion and deletion error probability, respectively, versus symbol index $k$ for $\alpha=0.2$, $\bar{T}^{\mathrm{symb}}=2$ ms, and $\text{SNR}_A=\text{SNR}_B = 3$ dB, i.e., the same parameters as in Figs.~\ref{Fig:PDF} and \ref{Fig:Eerror}. For the ML schemes,  the insertion and deletion error probabilities for Frameworks~1 and 2 are negligible and below $10^{-5}$ so that they cannot be seen in Figs.~\ref{Fig:InsDel} a) and b). Similarly, in Fig.~\ref{Fig:InsDel} b), the deletion error probability for the LF schemes is also below $10^{-5}$. For both frameworks, the insertion  error probability is higher for the LF scheme than for the TT scheme and highest for the PO scheme.  The deletion error probability is higher for the TT scheme than for the PO scheme which matches the order expected from Fig.~\ref{Fig:PDF}. We observe from Fig.~\ref{Fig:InsDel} that the insertion error probability increases with increasing symbol index $k$ whereas the deletion error probability decreases with increasing symbol index $k$.

\begin{figure}[t]
  \centering
    \resizebox{0.5\linewidth}{!}{\hspace{-2.7cm}
\psfragfig{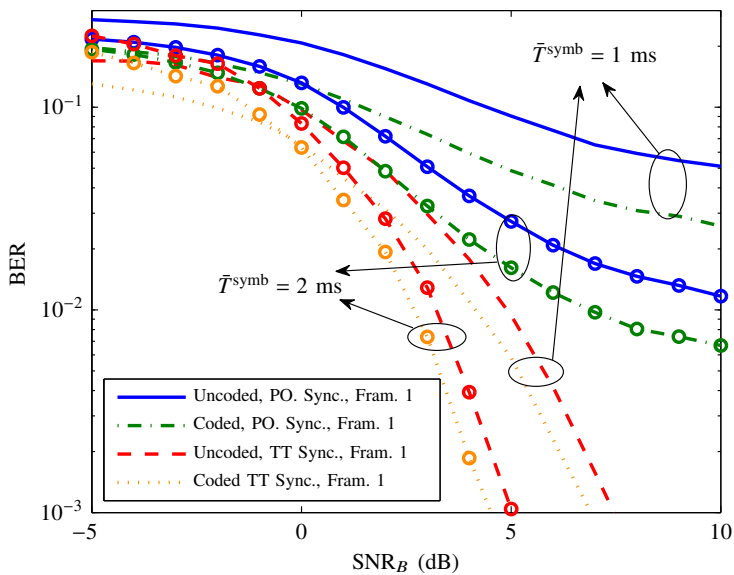}}
\caption{ BER versus $\text{SNR}_B$ for $\alpha=0.2$, $\text{SNR}_A=10$ dB, and $\bar{T}^{\mathrm{symb}}\in\{1,2\}$ ms. }
\label{Fig:BER_SNR_InsDel}  \vspace{-0.5cm}
\end{figure}

In order to cope with the performance loss resulting from insertion and deletion errors, we employ the code from \cite{InsDelMarkerCode} introduced in Section~IV-A. In particular, the last three symbols of each 10 symbols are the marker $100$, i.e., there is $30$ percent overhead,  and decoding is performed according to Table~II. For clarity of presentation, we include results only for the PO and TT schemes under Framework~1.  Note that the adopted codes were designed for correcting one insertion or one deletion error but not a substitution error, i.e., for the case when synchronization errors constitute the performance bottleneck  and not detection errors. Therefore, we assume a high SNR for the type-$A$ molecules, i.e., $\text{SNR}_A = 10$ dB, and vary the SNR of the type-$B$ molecules. In Fig.~\ref{Fig:BER_SNR_InsDel}, we show BER versus $\text{SNR}_B$ for $\alpha=0.2$, $\text{SNR}_A=10$ dB, and $\bar{T}^{\mathrm{symb}}\in\{1,2\}$ ms. We observe that the adopted insertion-deletion code improves the BER performance with respect to uncoded transmission for both the PO and TT schemes. Nevertheless, for a given synchronization scheme, the slopes of the curves for uncoded and coded transmission are identical. The limited effectiveness of the adopted code may be attributed to its inability to correct multiple deletion and insertion errors nor substitution errors. Hence, custom designed codes for the considered MC system are an interesting topic for future work. These codes could exploit e.g. knowledge regarding what type of error (insertion or deletion) is more likely.

\section{Conclusions}

In this paper, we considered an MC system where the transmitter is not equipped with an internal clock and is not restricted to emit the molecules with a constant release frequency. To enable symbol synchronization in this case, we proposed two frameworks which both employ two different types of molecule. In the first framework, one type of molecule is used for synchronization and the other type is used for data transmission whereas in the second framework, both types of molecule are used for joint symbol synchronization and data detection. We derived the optimal ML synchronization scheme  as a performance upper bound for each framework. As ML synchronization entails high complexity, we also developed three low-complexity synchronization schemes, namely the suboptimal LF, PO, and TT  schemes, for each framework. 

In the following, we summarize the main characteristics of the symbol synchronization schemes proposed in this paper. 

\begin{itemize}
\item The ML synchronization schemes achieve the best performance but also entail the highest complexity and require full knowledge of the MC channel.
\item  The LF synchronization schemes are computationally less complex than the ML schemes; however, they also require full knowledge of the MC channel. The LF schemes perform close to the ML schemes when the ISI is weak and the variation of the symbol duration is small. 
\item  The PO synchronization schemes are the simplest  among the considered schemes since they need the least a priori information about the channel and do not require  the constraint $T^{\max}\leq 2T^{\min}$,  which is needed for the ML and LF schemes; however, they may also introduce a significant performance loss.
\item  The TT synchronization schemes provide a favorable tradeoff between complexity and performance which makes them well suited for application in MC systems with limited computational capabilities.
\item Regarding the comparison of the two frameworks, our simulation results suggest that the ML scheme under Framework~2 outperforms the ML scheme under Framework~1; however, the suboptimal schemes under Framework~2  are all outperformed by the respective suboptimal schemes under Framework~1. The latter property can be attributed to the fact that an optimal detector was adopted for all synchronization schemes in Framework~1. We note that Framework~1 has the advantage that any modulation and detection scheme can be employed for  data transmission and straightforward generalization to the broadcast channel is possible without increasing the synchronization overhead. 
\end{itemize}

\bibliographystyle{IEEEtran}
\bibliography{Ref_03_08_2017}

\end{document}